\documentclass[aps,prx,reprint,onecolumn,12pt,eprint,raggedbottom,citeautoscript,superscriptaddress]{revtex4-2}

\usepackage{graphicx}
\usepackage{dcolumn}
\usepackage{xcolor}
\usepackage{soul}
\usepackage{bm}
\usepackage{amsmath}

\usepackage{ragged2e}   
\usepackage{subcaption}  

\usepackage{caption}
\usepackage[colorlinks=true,
            linkcolor=blue,
            citecolor=blue,
            urlcolor=blue]{hyperref}

\newcommand{\FUBMaffiliation}{\affiliation{Freie Universit\"at Berlin, Institute of Mathematics, Arnimallee 6, 14195 Berlin, Germany}}

\begin{document}

\title{Open Quantum System Embedded in an Optimized Reservoir of Molecules}


\author{Somayeh Ahmadkhani}
\email{s.ahmadkhani@fu-berlin.de}
\FUBMaffiliation
\author{Sara Panahian Jand}%
\FUBMaffiliation
\author{Luigi Delle Site}
\FUBMaffiliation

\begin{abstract}
Open-boundary Quantum Mechanics/Molecular Mechanics (QM/MM) simulation provides a direct way to test whether a quantum subsystem and its surrounding classical reservoir are physically consistent. Here we use this framework to replace a standard flexible TIP3P reservoir by a dipole-renormalized (DR) Amber-compatible three-site water model optimized against a full \textit{ab initio} trajectory of liquid water. The renormalization targets the molecular dipole distribution and adjusts the atomic charges, bonded, and Lennard-Jones interaction parameters through a stochastic optimization procedure. In the case of a large QM region ($\langle N\rangle\approx 440$), where the surface-to-volume ratio is small, the DR reservoir induces, as expected, Grand Canonical statistics for the quantum subsystem. In smaller subsystems ($\langle N\rangle\approx 190$), where the surface-to-volume condition is deliberately less favorable, the same reservoir shows substantially improved molecular consistency with the QM region relative to the non-renormalized flexible TIP3P model. These results show that quantum information at molecular dipole level is crucial for the microscopic physics of liquid water. When this concept is transferred into a classical reservoir, it can be used to extend the reliable regime of adaptive open-QM/MM water simulations even beyond the physically dictated Grand Canonical conditions.
\end{abstract}

\maketitle



\section{\label{intro} Introduction}

In molecular simulation, the open-boundary and adaptive-resolution methods provide a statistical-mechanical framework for studying subsystems that exchange molecules and energy with a reservoir \cite{physrep,Praprotnik2007AdResS,jcpessence,epjb}. Within this framework, the open-boundary QM/MM scheme \cite{cpc,advqmm} enhances the standard QM/MM framework by allowing molecules to exchange freely between the QM and MM regions. At thermodynamic equilibrium, and when the appropriate finite-size and boundary conditions are met, this exchange follows a Grand Canonical statistics \cite{physrep,jmp,mauricio,njp,advthsimeff,jpa, Reible2025}. The central condition for defining a Grand Canonical ensemble is a negligible surface-to-volume ratio \cite{huangbook,njp, prrjohn,prr,advphysx}: the interaction energy between the QM region and its surroundings must be small compared with the internal energy of the QM region itself. The corresponding thermodynamic relations at the adaptive coupling boundary have been analyzed explicitly for open molecular systems \cite{Gholami2021ThermoBoundary}. When this condition holds, the reservoir does not need to reproduce microscopic features of the quantum subsystem  explicitly; it must instead only impose the correct macroscopic thermodynamic state on the open subsystem \cite{njp}.

Under these conditions, the MM reservoir functions primarily as a thermodynamic bath. Agreement of the particle-number distribution $P(N)$, structural radial distribution functions, and other low-order observables between the open QM region and an equivalent subsystem extracted from a full \textit{ab initio} simulation then constitutes a test of Grand Canonical consistency of the open QM region \cite{PanahianJand2024OpenQM,physrep,prx}.
{At the microscopic level, the interaction energy across the open boundary can itself be treated as a fluctuating observable whose mean and variance can be expressed in terms of the one-particle density and structural correlation functions coupling the subsystem to its reservoir {\cite{rahbar2026endpoint}}.}

When the QM region is made smaller, the surface-to-volume ratio grows and the QM/MM interface becomes dynamically significant. Molecules cross the boundary at every time step, and if the classical description of MM molecules is incompatible with the quantum molecular description, errors accumulate at the interface — producing unphysical density variations, distorted particle-exchange statistics, and artifacts in electronic-structure observables. This challenge is well recognized in adaptive QM/MM methods, where QM and MM assignments change throughout the dynamics \cite{aqmmm2,Zheng2016AdaptiveQMMM,Mones2015AdBF}. The adaptive buffered-force scheme implemented in CP2K and AMBER addressed such a problem by augmenting the QM region with a buffer zone whose size is chosen from force-convergence criteria, yielding stable adaptive dynamics for liquid water and reaction benchmarks \cite{Mones2015AdBF}. The present work, instead of a practical numerical engineering approach, addresses the question from a statistical mechanics point of view by focusing on the physical consistency of the molecular reservoir itself.

To this end, we renormalize a flexible Amber-style three-site water model directly against a full \textit{ab initio} trajectory of liquid water. The renormalization is formulated as a stochastic $\arg\min$ problem in which the molecular dipole distribution serves as the primary electronic target \cite{argmin}. The molecular dipole distribution for liquid water is considered one of the most relevant quantum signatures of such a system \cite{pasquarello2003dynamical,shi2021condensed}. The optimized parameters define the MM reservoir for two main open-QM/MM validation cases: a large region with $\langle N\rangle\approx 440$, where Grand Canonical behavior is expected, and a smaller region with $\langle N\rangle\approx 190$, where the same reservoir is probed under the unfavorable condition of sizable QM-MM interface effects. In contrast to buffered-force approaches, which modify the QM/MM boundary treatment, our strategy fixes the open-QM/MM protocol and the Amber functional form while improving the quality of the classical environment. Throughout, comparison with equivalent subsystems extracted from the full \textit{ab initio} trajectory provides the quantum reference. In addition to the two principal validation cases, an extreme small-region
calculation containing on average $\langle N\rangle\approx 50$ water molecules is reported in the Appendix~\ref{app:qm50_results}. It obviously probes an extreme regime: at this size, the boundary interaction is almost dominant relative to the internal contribution of the QM subsystem. The surface-to-volume ratio condition required for a rigorous Grand Canonical description is therefore strongly violated.
The energy-ratio analysis of the $\langle N\rangle\approx50$ case further confirms this interpretation: the boundary contribution is no longer negligible relative to the internal QM contribution. 
In this case, the calculation tests the level of very local compatibility of the optimized reservoir with the quantum description when even nearly every QM molecule is influenced by the adaptive interface.

\section{\label{application}Implementation of the Protocol}

\begin{figure}
 \centering
    \includegraphics[width=0.60\textwidth]{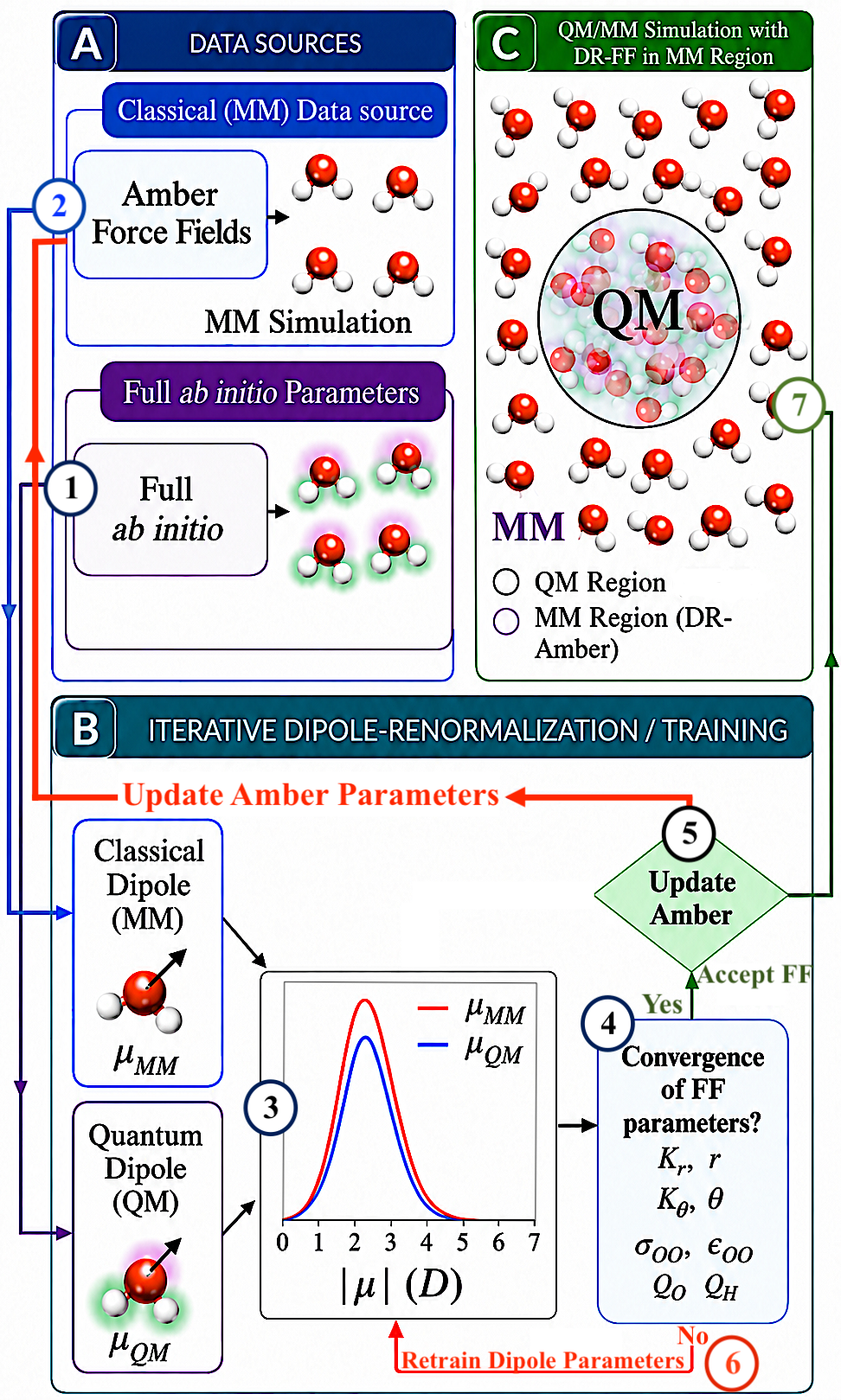}
    \caption{ \justifying
    {Schematic representation of the dipole-renormalized open-QM/MM protocol. (A) Quantum and classical data sources. (B) Coupled iterative adjustment of dipole-related and remaining force-field parameters. The first decision checks whether the dipole-distribution match is retained after adjusting the remaining force-field parameters: ``Yes'' (step~5) leads to an updated classical simulation and reassessment of parameter convergence, whereas ``No'' (step~6) returns to dipole matching through small parameter perturbations. If the parameters have not converged, the optimization cycle is repeated. (C) Once the parameters converge while preserving the dipole-distribution match, the final Amber set of parameters is used in adaptive open-QM/MM production simulations (step~7). Circled numbers correspond to the steps described in the text.}
    }
    \label{fig:protocol_cartoon}
\end{figure}

The protocol proceeds in two stages: construction of the optimized MM reservoir, and validation through open QM/MM simulation at two main QM-region sizes. A schematic overview of the workflow is shown in Fig.~\ref{fig:protocol_cartoon}.

\subsection*{Reservoir construction}

Following the workflow illustrated in Fig.~\ref{fig:protocol_cartoon}, a full ab initio trajectory of liquid water serves as the sole source of target observables.
The MM reservoir is a flexible Amber-style three-site water model whose independent parameters are the oxygen charge $Q_{\rm O}$, the hydrogen charge fixed by neutrality $Q_{\rm H}=-Q_{\rm O}/2$, the O--H equilibrium distance and bond-stretching force constant $(r^0_{\rm OH},k_r)$, the H--O--H equilibrium angle and angular force constant $(\theta^0_{\rm HOH},k_\theta)$, and the oxygen--oxygen Lennard-Jones parameters $(\sigma_{\rm OO},\epsilon_{\rm OO})$. These parameters are optimized against the \textit{ab initio} reference.
This step belongs to the broader class of force-field fitting or parameter training procedures, where parameters in a fixed analytical potential are adjusted to reproduce reference data; established examples include force matching, multiscale coarse graining, relative-entropy optimization, and automated force-field optimization \cite{ercolessi1994force,izvekov2005multiscale,shell2008relativeentropy,wang2013systematicwater,wang2014forcebalance}. In the present work, "training" refers only to this bounded parameter-fitting step. The only electronic information transferred into the reservoir is the molecular dipole distribution; no higher-multipole targets are used. This choice is physically motivated because molecular dipoles obtained from the electronic charge density, or equivalently from maximally localized Wannier centers, provide a direct low-order descriptor of condensed-phase polarization in quantum liquid water: first-principles studies showed that the liquid-state dipole is enhanced relative to the gas phase, that dynamical molecular dipoles control dielectric and infrared response, and that dipole fluctuations correlate with local hydrogen-bond structure \cite{wannier1,wannier2,silvestrelli1999water,pasquarello2003dynamical,sharma2005intermolecular,mcgrath2007spatial}. At the same time, quadrupolar observables and molecular multipoles are known to provide sensitive information on local electronic structure and are often emphasized as important ingredients for models that aim to retain quantum character in molecular simulations \cite{shi2021condensed,keeler2019high,xie2024electric}. The dipole target is therefore a deliberate compromise rather than a claim that quadrupoles are irrelevant: a three-site point-charge model has sufficient degrees of freedom to match the charge centroid of a water molecule, but not its quadrupole or higher moments.

The \textit{ab initio} target dipoles are obtained from localized Wannier-center data \cite{wannier1,wannier2}, not from the three-site MM expression. By contrast, the molecular dipole of a candidate three-site water molecule in the MM model is
\begin{equation}
    \boldsymbol{\mu} = Q_{\rm O}\mathbf{r}_{\rm O} + Q_{\rm H}\mathbf{r}_{\rm H_1}+Q_{\rm H}\mathbf{r}_{\rm H_2},
    \label{eq:mm_dipole}
\end{equation}
with magnitudes converted from $e\,\text{\AA}$ to Debye in both cases. The objective function minimized during optimization is
\begin{equation}
    \mathcal{L}=W_1\!\left[P_{\rm MM}(\mu),P_{\rm target}(\mu)\right]
    +2\left|\langle\mu\rangle_{\rm MM}-\langle\mu\rangle_{\rm target}\right|
    +\left|\sigma_{\mu,\rm MM}-\sigma_{\mu,\rm target}\right|,
\end{equation}
where $P_{\rm target}(\mu)$ is the Wannier-center \textit{ab initio} dipole distribution, $P_{\rm MM}(\mu)$ is the classical distribution computed from Eq.~\eqref{eq:mm_dipole}, {$\sigma_{\mu}$ denotes the standard deviation (width) of the molecular-dipole distribution}, and $W_1$ is the one-dimensional Wasserstein distance, evaluated with SciPy \cite{scipy}. The loss thus penalizes errors in the full shape of the dipole distribution, its mean, and its width simultaneously. In practice, the dipole is not modified directly; it is the target observable used to optimize the charges, bonded parameters, and Lennard-Jones terms of the three-site force field.

{The term ``dipole-renormalized'' refers to the quantum-derived observable used as the optimization target, rather than to an optimization of the partial charges alone. A preliminary adjustment of the dipole-related parameters improved the dipole distribution but did not provide stable convergence in the full MM simulation. This is because the molecular dipole also depends on the instantaneous bond lengths and molecular angle, while the Lennard--Jones interactions influence the surrounding molecular environment. Consequently, the charges, bonded terms, and Lennard--Jones parameters were optimized through a coupled iterative procedure. The improvement obtained with the final DR reservoir should therefore be attributed to the complete dipole-targeted reparameterization and not exclusively to the change in partial charges.}

{Below we will provide a {step-by-step} description of the optimization procedure having as a starting point the MM model of flexible-TIP3P \cite{price2004fTIP3P}. The procedure is also pictorially illustrated in Fig.~\ref{fig:protocol_cartoon}:}

\begin{enumerate}
\setcounter{enumi}{-1}
\item {{The classical MM simulations used during the dipole-renormalization procedure contained 2960 water molecules, identical to the full {\it ab initio} reference system. The usual equilibration protocol is applied by running an NPT simulation with CP2K~\cite{bib:cp2k}. The average box size with a side length of approximately 42.24~\AA~obtained from the NPT equilibration is used for both the full {\it ab initio} and full classical NVT simulations.}
}

\item {A full {\it ab initio} simulation is performed to obtain the reference molecular dipole distribution.}

\item {In parallel, a classical simulation using the flexible-TIP3P parameters is performed to obtain the classical molecular dipole distribution, which is compared with the reference distribution described above.}

\item {Based on the comparison between the {\it ab initio} and classical dipole distributions, modifications to the dipole-related parameters ($r$, $\theta$, and $q$) are proposed. The equilibrium bond length $r$, bond angle $\theta$, and oxygen charge $q$ are varied using step sizes of {$\delta_r=0.005~\AA$, $\delta_\theta=0.5^\circ$, and $\delta_q=0.02\,e$, where $e=1.602176634\times10^{-19}~\mathrm{C}$ is the elementary charge.}, respectively. A short classical simulation is then performed, and the corresponding dipole distribution is compared with the reference distribution.
{This process is repeated until the maximum absolute difference between the corresponding normalized histogram-bin probabilities is smaller than 0.001. This threshold is dimensionless.}
}
\item {
Because, in a multi-parameter problem, the values of $r$, $\theta$, and $q$ obtained above may not necessarily correspond to a global minimum, we perform an additional cross-check. The modified force field containing the new values of $r$, $\theta$, and $q$ is used, while the remaining force-field parameters,
{namely $K_r$, $K_\theta$, $\sigma_{\mathrm{OO}}$, and
$\epsilon_{\mathrm{OO}}$,} are also allowed to vary through small perturbations. {The perturbation sizes are
$\delta K_r=20.0~\mathrm{kcal\,mol^{-1}\,\AA^{-2}}$,
$\delta K_\theta=5.0~\mathrm{kcal\,mol^{-1}\,rad^{-2}}$,
$\delta\sigma_{\mathrm{OO}}=0.03~\AA$, and
$\delta\epsilon_{\mathrm{OO}}=0.01~\mathrm{kcal\,mol^{-1}}$.}
The variation of {these four parameters} is governed by the condition that the {trajectory-averaged modified forces} match the corresponding {\it ab initio} forces {with a maximum absolute force difference smaller than
$0.001~\mathrm{Hartree}\,\AA^{-1}$.} Using the updated force field, we then perform a short classical simulation and calculate the resulting classical dipole distribution.
}

\item {If the agreement between the classical dipole distribution and the reference distribution is retained/achieved (the ``Yes'' branch in Fig.~\ref{fig:protocol_cartoon}), we accept the new Amber set of parameters.}

\item {If the agreement is not achieved (the ``No'' branch in Fig.~\ref{fig:protocol_cartoon}), we modify the dipole-related parameters again and repeat the procedure until the required agreement is achieved.}

\item {Once successive classical simulation and parameter-assessment cycles yield converged parameters while preserving the dipole-distribution match, generate the final Amber set of parameters and use the optimized reservoir in the adaptive open-QM/MM simulations.}
\end{enumerate}

{Trial updates that increase the objective function are rejected, and the current best parameter set is retained. The convergence of the molecular dipole and the retained force-field parameters is reported in Appendix~\ref{app:optimization_convergence}.}

\subsection*{Quantification of QM/MM interface effects:
The {surface-to-volume ratio}}

To monitor how strongly the QM--MM boundary contributes to the total energy, we decompose the instantaneous energy as
\begin{equation}
    E_{\rm QMMM}(t)=E_{\rm QM}(t)+E_{\rm MM}(t)+E_{\rm int}(t),
\end{equation}
where $E_{\rm QM}(t)=N_{\rm QM}(t)\langle\epsilon_{\rm QM}\rangle$ and $E_{\rm MM}(t)=N_{\rm MM}(t)\langle\epsilon_{\rm MM}\rangle$ are estimated from the corresponding full-resolution averages, giving $E_{\rm int}(t)=E_{\rm QMMM}(t)-E_{\rm QM}(t)-E_{\rm MM}(t)$. The surface-to-volume diagnostic is then
\begin{equation}
    R(t)=\left|\frac{|E_{\rm QM}(t)|-|E_{\rm int}(t)|}{E_{\rm QM}(t)}\right|.
\end{equation}
When $R(t)$ approaches unity, the internal QM energy dominates the boundary interaction and the surface-to-volume condition is satisfied. Significant departures from unity signal that molecular-level reservoir consistency is being tested. In statistical mechanics $R(t)$ is called the surface-to-volume criterion and is used for the derivation of the formal expression of the partition function of the Grand Canonical ensemble (see e.g. \cite{huangbook}).

\section{\label{tecdet} Technical Details}

Where possible, we follow the protocol of Jand \textit{et al.}~\cite{PanahianJand2024OpenQM} and explicitly note where our setup differs.

\subsection{System preparation and \textit{ab initio} setup}

Liquid water configurations were pre-equilibrated classically using the flexible TIP3P (fTIP3P) water model \cite{price2004fTIP3P}. Initial simulation boxes were generated with AmberTools \cite{case2023ambertools}, followed by short energy minimization and NPT/NVT runs to reach the target density at 300~K and 1~bar.

These equilibrated structures were used as starting configurations for DFT-based \textit{ab initio} molecular dynamics simulations \cite{abinit} in CP2K \cite{bib:cp2k}. The electronic structure was treated within Kohn--Sham DFT using the PBE exchange--correlation functional \cite{perdew1996pbe}, Goedecker--Teter--Hutter pseudopotentials \cite{goedecker1996gth}, and a DZVP-MOLOPT-SR-GTH Gaussian basis set \cite{vandevondele2007basis} within the Gaussian and Plane Waves scheme, following the CP2K/Quickstep protocol \cite{vandevondele2005quickstep}. Unless stated otherwise, a plane-wave cutoff of 280~Ry and a relative cutoff of 30~Ry were used, following Ref.~\cite{PanahianJand2024OpenQM} and the technical choices of Ref.~\cite{shao2003curvy}. Dispersion interactions were included through the DFT-D3 scheme \cite{grimme2010dft}. All self-consistent-field calculations used the orbital-transformation method with convergence thresholds between $10^{-6}$ and $10^{-5}$~Ha. Periodic boundary conditions and Ewald summation were applied throughout.

\subsection{Full \textit{ab initio} molecular dynamics reference}

The \textit{ab initio} reference simulations were performed in the NVT ensemble using CP2K. 
{The reference system contained 2960 water molecules. The initial cubic box of side length 44.5~\AA{} was equilibrated in the NPT ensemble, yielding a final side length of approximately 42.24~\AA{}.} 
A time step of 1.0~fs was used, and trajectories were propagated for 20~ps after equilibration. Temperature was controlled at 300~K with an adaptive Langevin thermostat \cite{jones_langevin}, and center-of-mass translation and total angular momentum were removed at the start to prevent drift. Maximally localized Wannier centers were computed every five MD steps as implemented in CP2K \cite{bib:cp2k,wannier1}; following established liquid-water Wannier analyses \cite{wannier2,silvestrelli1999water}, they are used here exclusively to extract molecular dipole targets for the reservoir optimization.

\subsection{Open QM/MM simulations}

Open QM/MM simulations were carried out using the adaptive open-QM/MM implementation in CP2K at 300~K \cite{PanahianJand2024OpenQM,bib:cp2k,Cp2k_parinello}. The total system contained 4320 water molecules in a cubic box of side length 50.304~\AA, matching the setup of Ref.~\cite{PanahianJand2024OpenQM}. A spherical QM region was placed at the center of the box. The main DR validation uses two QM-region radii: 14.5~\AA, containing on average $\langle N \rangle \approx 440$ water molecules (the large-region, near-Grand-Canonical test), and 11.21~\AA, containing on average $\langle N \rangle \approx 190$ water molecules (the smaller-region, strong-interface test). An additional spherical QM region containing on average $\langle N\rangle\approx 50$ water molecules, with an approximate radius of $7.0~\text{\AA}$, was also simulated. Because the interaction with the surrounding reservoir is no longer small compared with the internal QM contribution at this size, this system does not satisfy the surface-to-volume condition required for a Grand Canonical description. Its results are therefore presented in Appendix~\ref{app:qm50_results} as an extreme molecular-compatibility test rather than as a thermodynamic open-system
validation.

Molecules were reclassified as QM or MM at every time step based on their instantaneous positions, following the adaptive reassignment logic used in buffered-force methods to allow exchange with the high-level region without artificial density drift \cite{aqmmm2,Zheng2016AdaptiveQMMM,Mones2015AdBF}. The QM region used the PBE-D3 setup specified above, and the MM region used the optimized Amber reservoir. All QM/MM runs were performed for 20~ps in the NVT ensemble with a 1.0~fs time step; center-of-mass motion was fixed, an adaptive Langevin thermostat \cite{jones_langevin,Mones2015AdBF} supported the adaptive partitioning, and Wannier centers were localized every five steps to obtain QM-region molecular dipoles.
{For direct comparison,
{Figs.~\ref{fig:pn_440}--\ref{fig:rdf_decomp_190}}
also include the corresponding open-QM/MM results obtained with the flexible TIP3P reservoir in {Ref.~\cite{PanahianJand2024OpenQM}}. A direct TIP3P dipole comparison is available only for the {$\langle N\rangle\approx440$} region.}

\section{\label{results}Results}

Following the protocol for defining a physically meaningful open quantum system introduced in Ref.~\citenum{PanahianJand2024OpenQM} we compare each open QM region with the corresponding subsystem extracted from the full {\it ab initio} trajectory using four classes of observables: the particle-number distribution $P(N)$; the oxygen--oxygen and oxygen--hydrogen radial distribution functions $g_{\rm OO}(r)$ and $g_{\rm OH}(r)$; the Wannier-resolved structural and molecular-dipole observables; and the energetic ratio $R(t)$, which measures the relative importance of the QM/MM boundary.

For QM regions satisfying the negligible surface-to-volume ratio condition, the agreement of those quantities calculated in the QM region with the full {\it ab initio} reference provides a relevant test regarding the Grand Canonical particle exchange statistics. For smaller QM regions instead QM/MM boundary-effects are sizable and the diagnostic will show whether one has  conditions equilibrium. Equilibrium is in this case a necessary condition for determine the level of microscopic molecular compatibility between the QM and MM models.

\begin{figure}[!ht]
    \centering
    \includegraphics[width=0.99\linewidth]{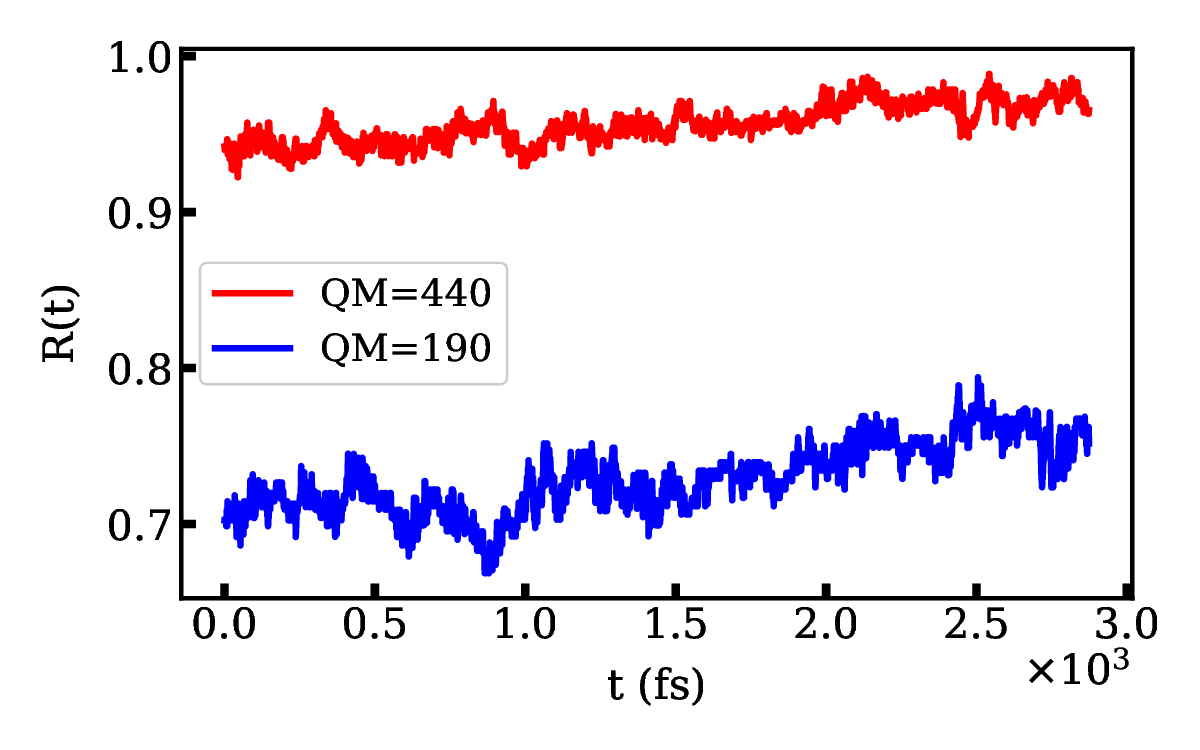}
    \caption{\justifying Time evolution of the energy ratio $R(t)$ for the 440-molecule large QM-region structure ($\langle N \rangle \approx 440$) and the 190-molecule small QM-region structure ($\langle N \rangle \approx 190$). Values near unity indicate that the internal QM energy dominates the boundary interaction; departures signal that the surface-to-volume condition is violated and that molecular-level reservoir consistency is being tested. The corresponding fluctuation ranges are $\Delta R=0.048$ and $0.100$, respectively.}
    \label{fig:energy_440_190}
\end{figure}

\subsection{\label{440} Large QM region: $\langle N \rangle \approx 440$ molecules}

\begin{figure}[!ht]
    \centering
        \includegraphics[width=0.99\linewidth]{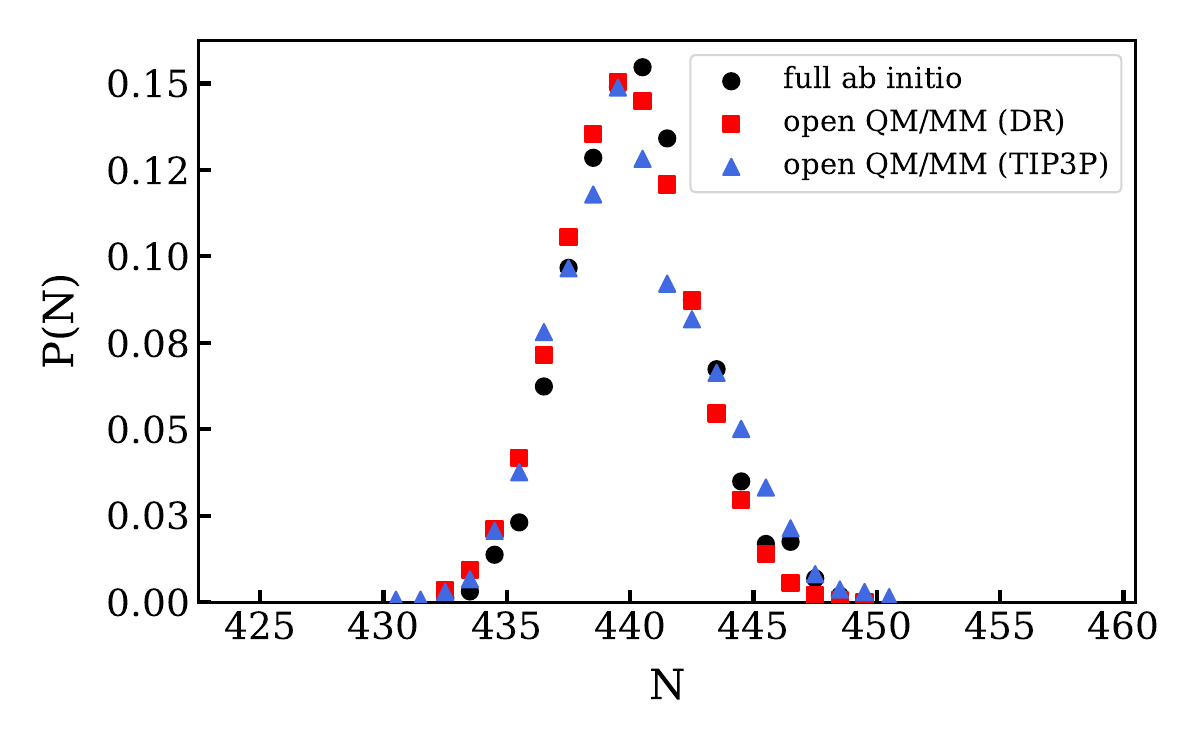}
    \caption{\justifying Particle-number distribution $P(N)$ for the 440-molecule large QM-region structure. Agreement with the full \textit{ab initio} subsystem confirms that the optimized reservoir reproduces Grand Canonical exchange when the surface-to-volume condition is satisfied. {For comparison the flexible-TIP3P data, corresponding to Fig.~5 of Ref.~\cite{PanahianJand2024OpenQM}, are reported. It must be noted that $N$ of Ref. ~\cite{PanahianJand2024OpenQM} has been shifted to match with the two other plots. The discrepancy w.r.t. $\langle N\rangle$ of reference is less than $5\%$ though, which in the previous work was been classified as acceptable. One can immediately notice that the MM force field of this work, by refining the molecular dipole distribution w.r.t. the {\it ab initio} reference, does not show the discrepancy for $\langle N\rangle$. The data of Ref.~\cite{PanahianJand2024OpenQM} are used with permission under the Creative Commons Attribution--NonCommercial--NoDerivatives (CC BY-NC-ND 4.0) license.}}
    \label{fig:pn_440}
\end{figure}

\begin{figure}[!ht]
    \centering
    \begin{subfigure}[b]{0.49\linewidth}
        \centering
        \includegraphics[width=\linewidth]{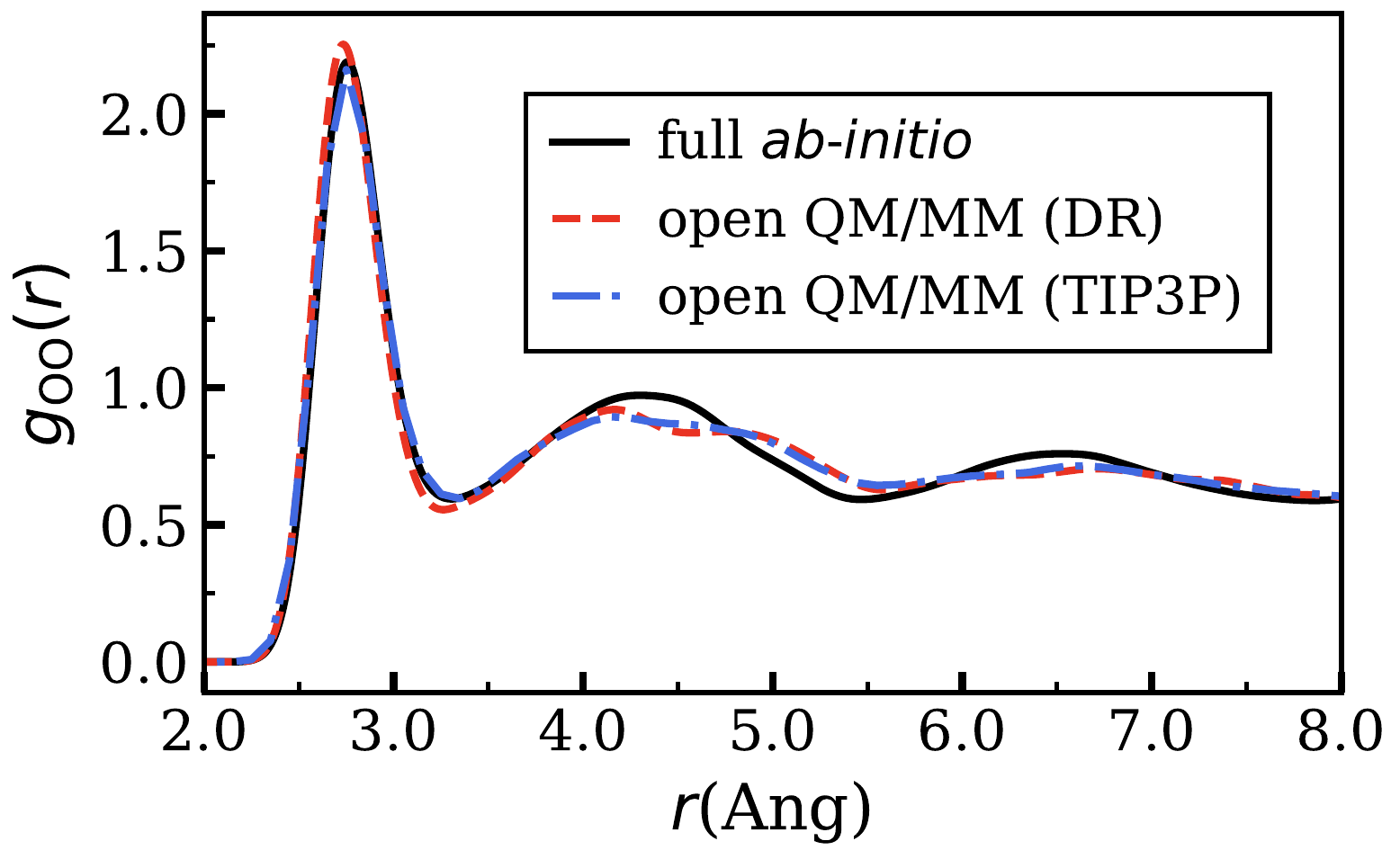}
        \caption{Oxygen--oxygen (OO) RDF}
    \end{subfigure}
    \hfill
    \begin{subfigure}[b]{0.49\linewidth}
        \centering
        \includegraphics[width=\linewidth]{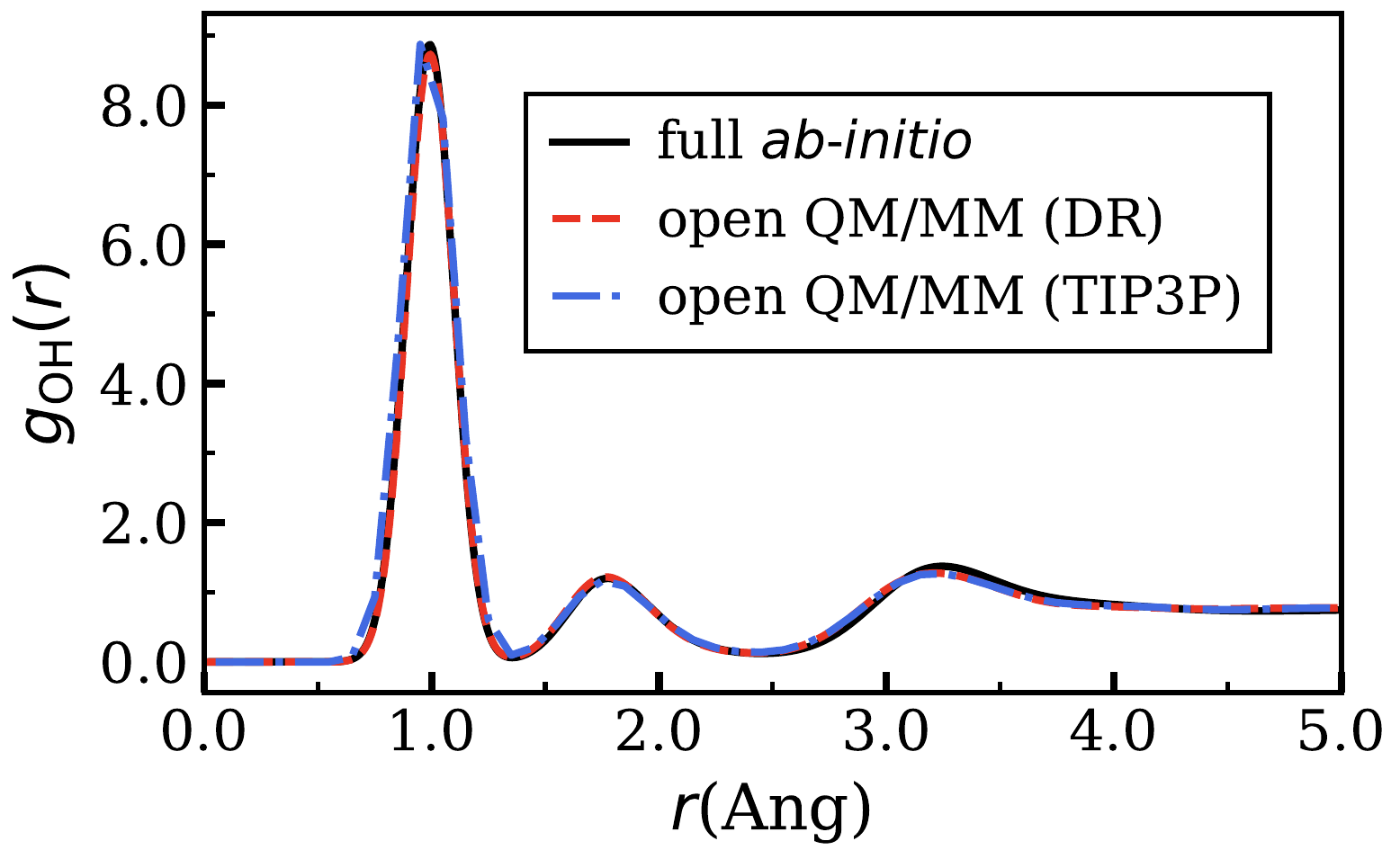}
        \caption{Oxygen--hydrogen (OH) RDF}
    \end{subfigure}
    \caption{\justifying
    Structural radial distribution functions for the 440-molecule large QM-region structure
    ($\langle N \rangle \approx 440$). Agreement with the full {\it ab initio}
    subsystem verifies that the optimized reservoir preserves the oxygen--hydrogen
    liquid structure in the large-region limit. The flexible-TIP3P data,
    corresponding to Figs.~3 and 4 of Ref.~\cite{PanahianJand2024OpenQM},
    are used with permission under the Creative Commons
    Attribution--NonCommercial--NoDerivatives (CC BY-NC-ND 4.0) license.
    }
    \label{fig:rdf_oh_400}
\end{figure}

For the 440-molecule QM region, the surface-to-volume ratio is small and the reservoir is expected to act primarily as a thermodynamic bath. The energy-ratio diagnostic in Fig.~\ref{fig:energy_440_190} confirms this: $R(t)$ remains close to unity throughout the simulation, with a small fluctuation range of $\Delta R=0.048$, indicating that the QM--MM boundary contributes negligibly to the total energy. In this regime, the decisive test is whether the optimized reservoir reproduces the correct thermodynamic exchange, and the particle-number distribution (according to the open system protocol of Ref.\citenum{PanahianJand2024OpenQM}. Fig.~\ref{fig:pn_440} shows that $P(N)$ is in good agreement with the equivalent subsystem extracted from the full {\it ab initio} trajectory and the same holds for the structural RDFs in Fig.~\ref{fig:rdf_oh_400}, i.e.,  the oxygen--oxygen and oxygen--hydrogen channels remain consistent with the reference data from full {\it ab initio} simulations.

\begin{figure}[!ht]
    \centering
    \begin{subfigure}[b]{0.47\linewidth}
        \centering
        \includegraphics[width=\linewidth]{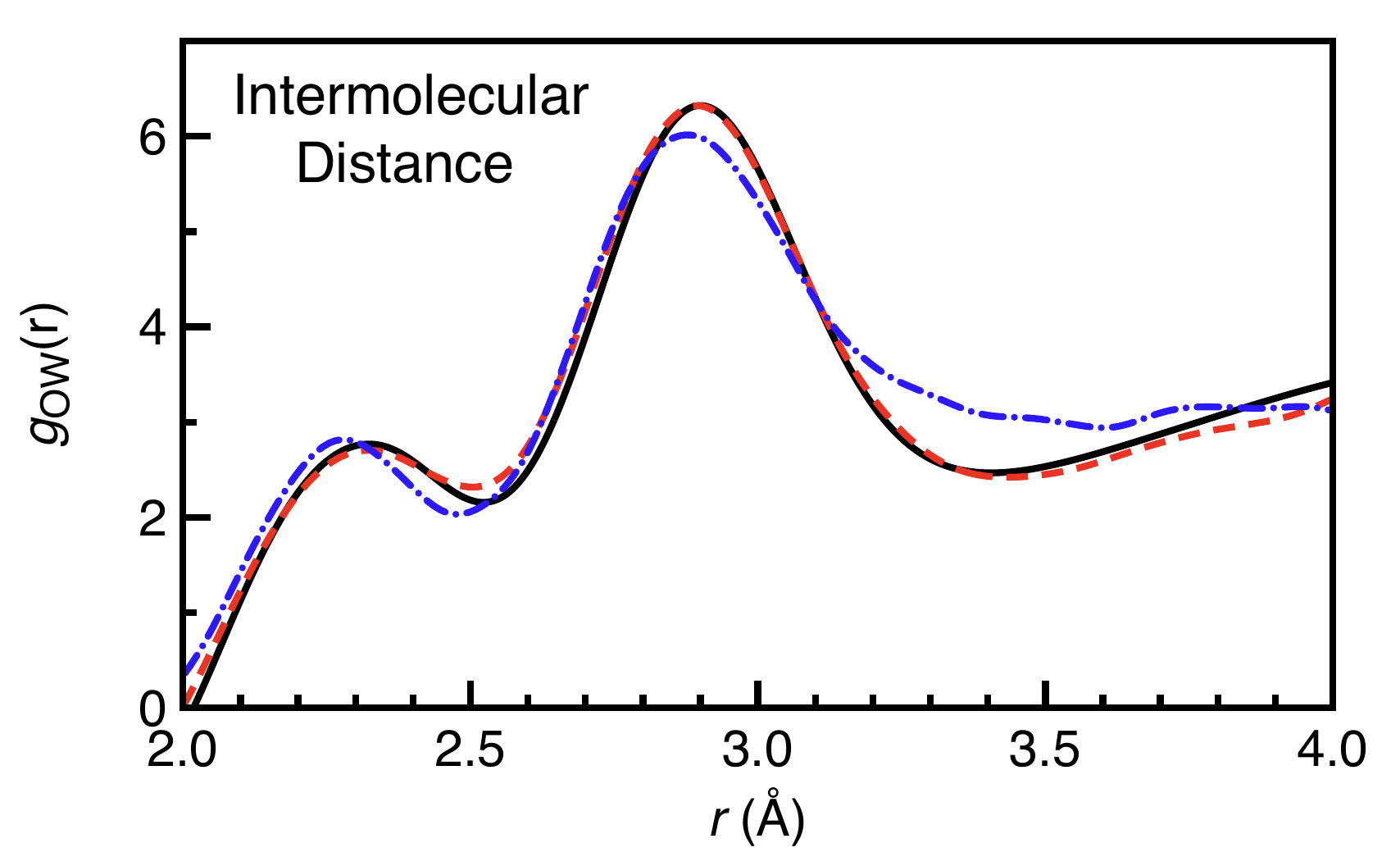}
        \caption{Intermolecular oxygen--Wannier (inter-OW)}
    \end{subfigure}
    \hfill
    \begin{subfigure}[b]{0.47\linewidth}
        \centering
        \includegraphics[width=\linewidth]{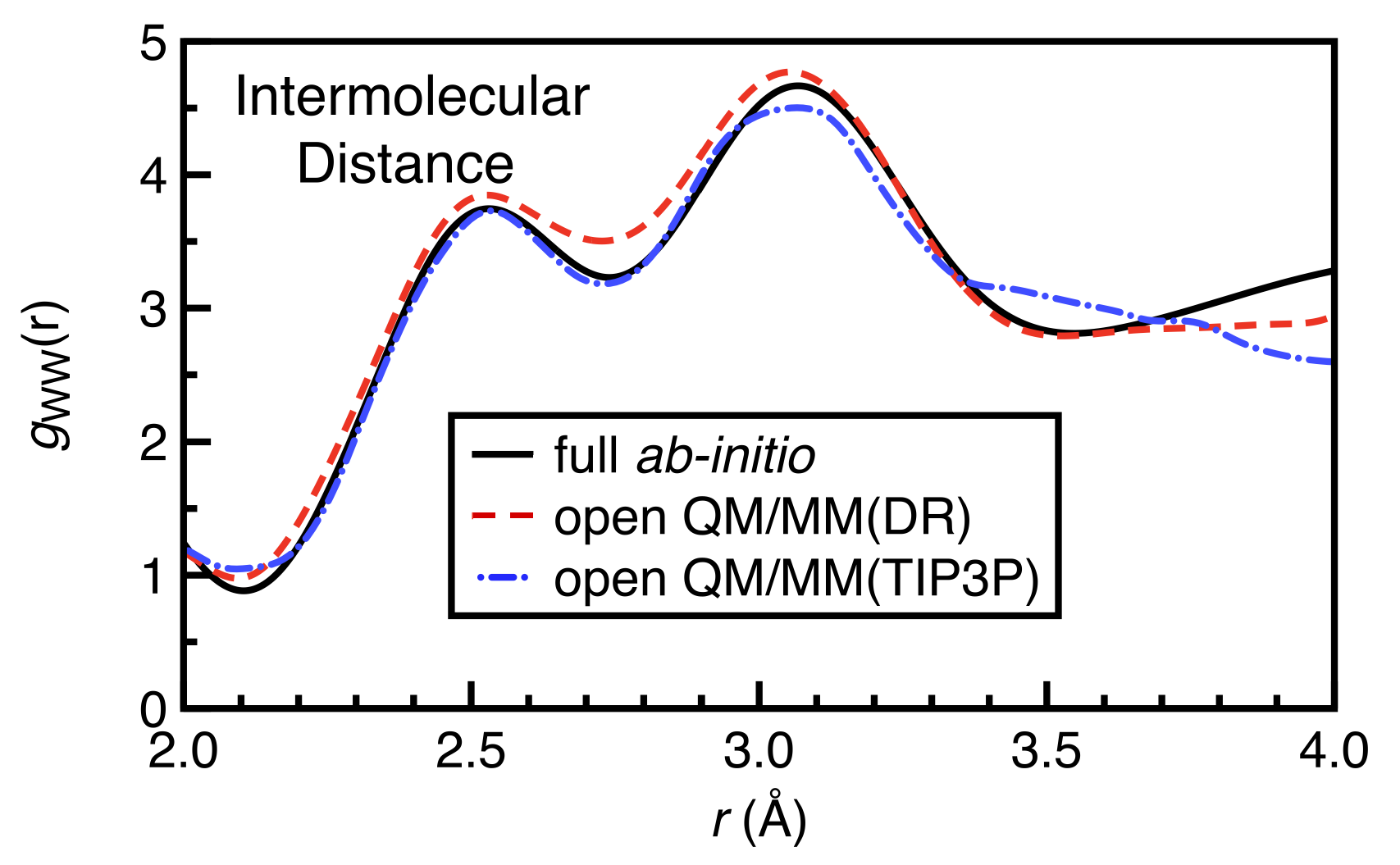}
        \caption{Intermolecular Wannier--Wannier (inter-WW)}
    \end{subfigure}
    \vspace{0.2cm}
    \begin{subfigure}[b]{0.47\linewidth}
        \centering
        \includegraphics[width=\linewidth]{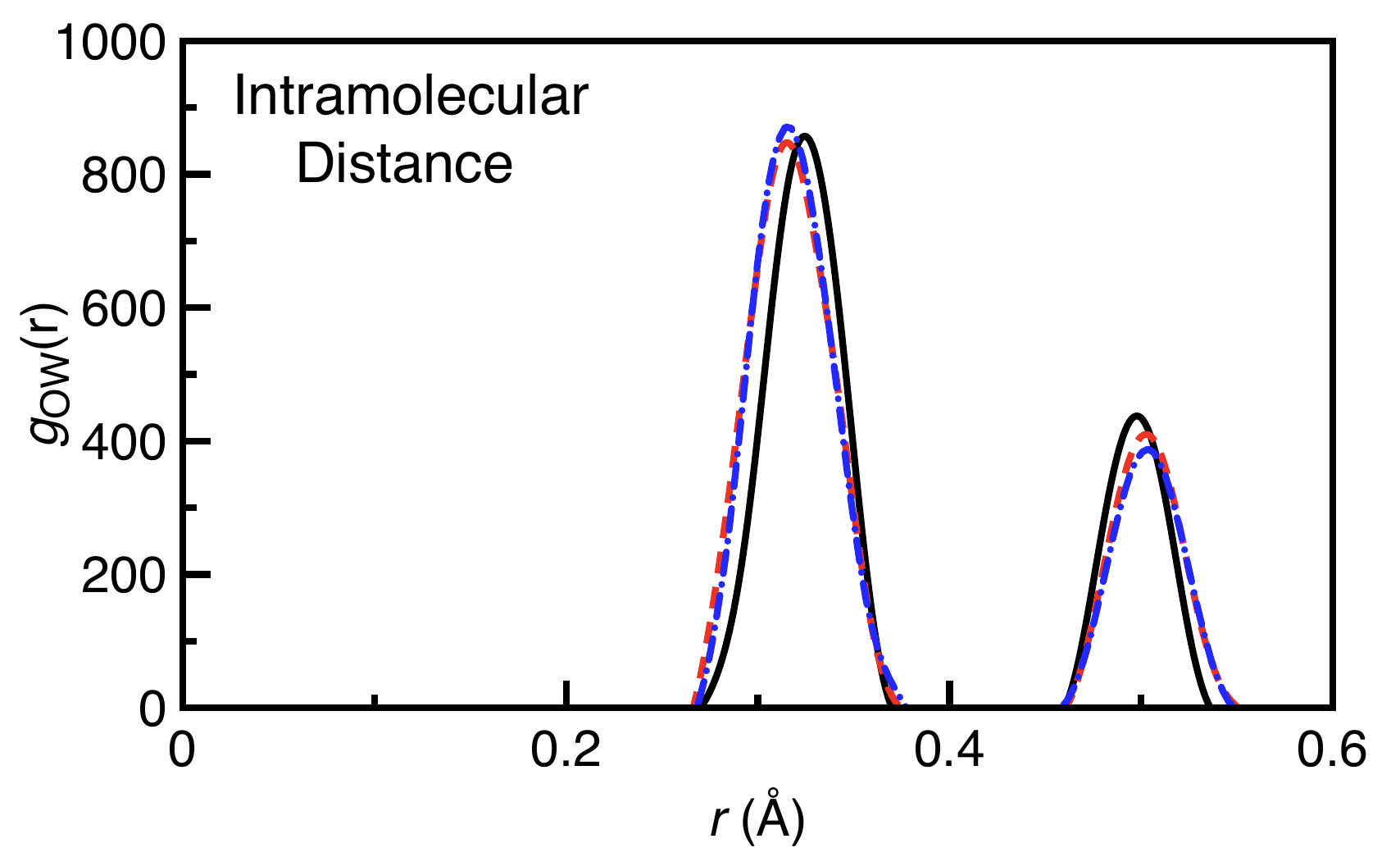}
        \caption{Intramolecular oxygen--Wannier (intra-OW)}
    \end{subfigure}
    \hfill
    \begin{subfigure}[b]{0.47\linewidth}
        \centering
        \includegraphics[width=\linewidth]{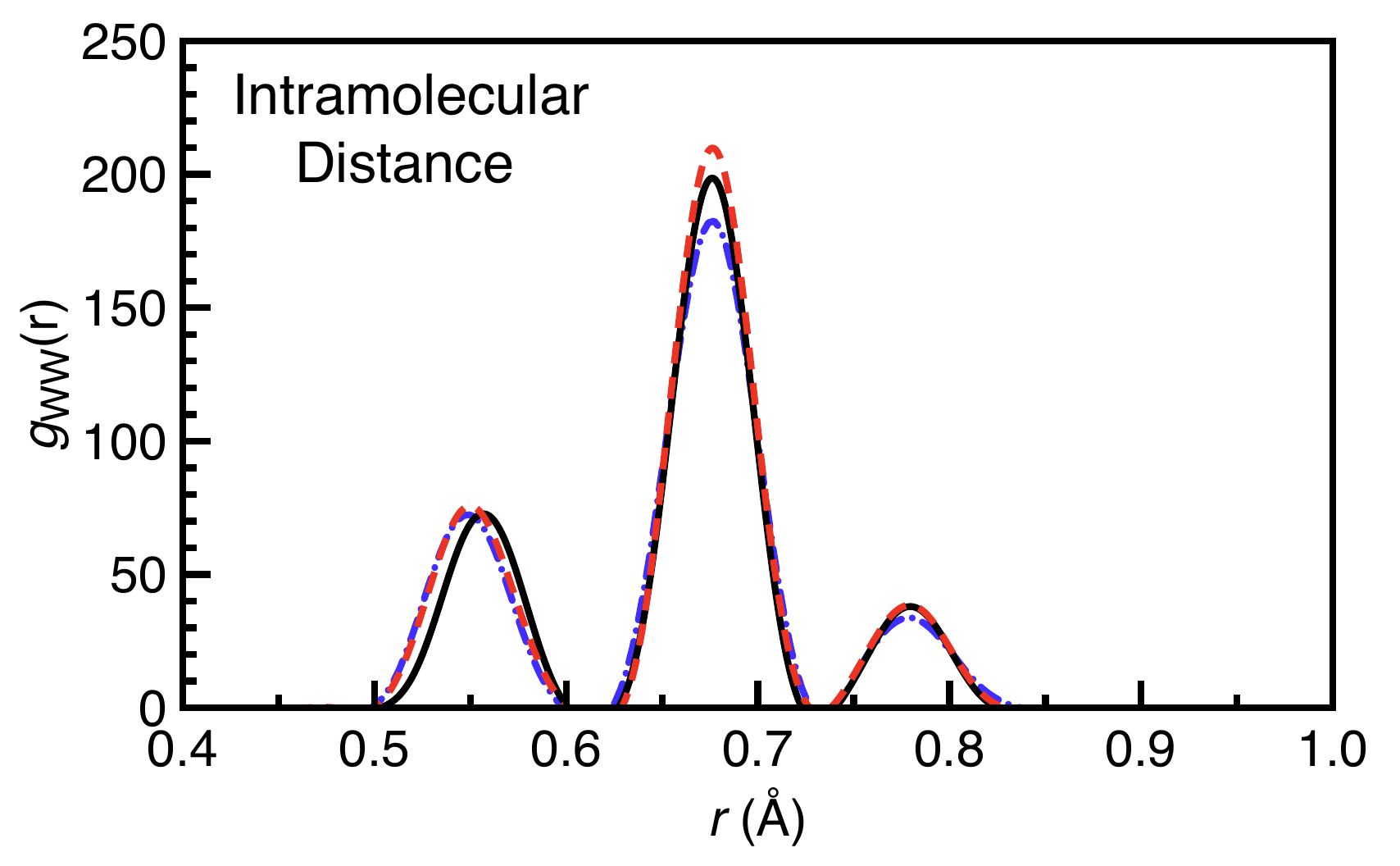}
        \caption{Intramolecular Wannier--Wannier (intra-WW)}
    \end{subfigure}
    \caption{\justifying
    Intermolecular (top) and intramolecular (bottom) decomposition of the
    Wannier-resolved RDFs for the 440-molecule large QM-region structure.
    This decomposition separately resolves the local molecular geometry and
    the surrounding liquid organization. The flexible-TIP3P data,
    corresponding to Figs.~7, 8, 10, and 11 of
    Ref.~\cite{PanahianJand2024OpenQM}, are used with permission under the
    Creative Commons Attribution--NonCommercial--NoDerivatives
    (CC BY-NC-ND 4.0) license.
    }
    \label{fig:rdf_decomp_400}
\end{figure}

\begin{figure}[!ht]
    \centering
    \includegraphics[width=\linewidth]{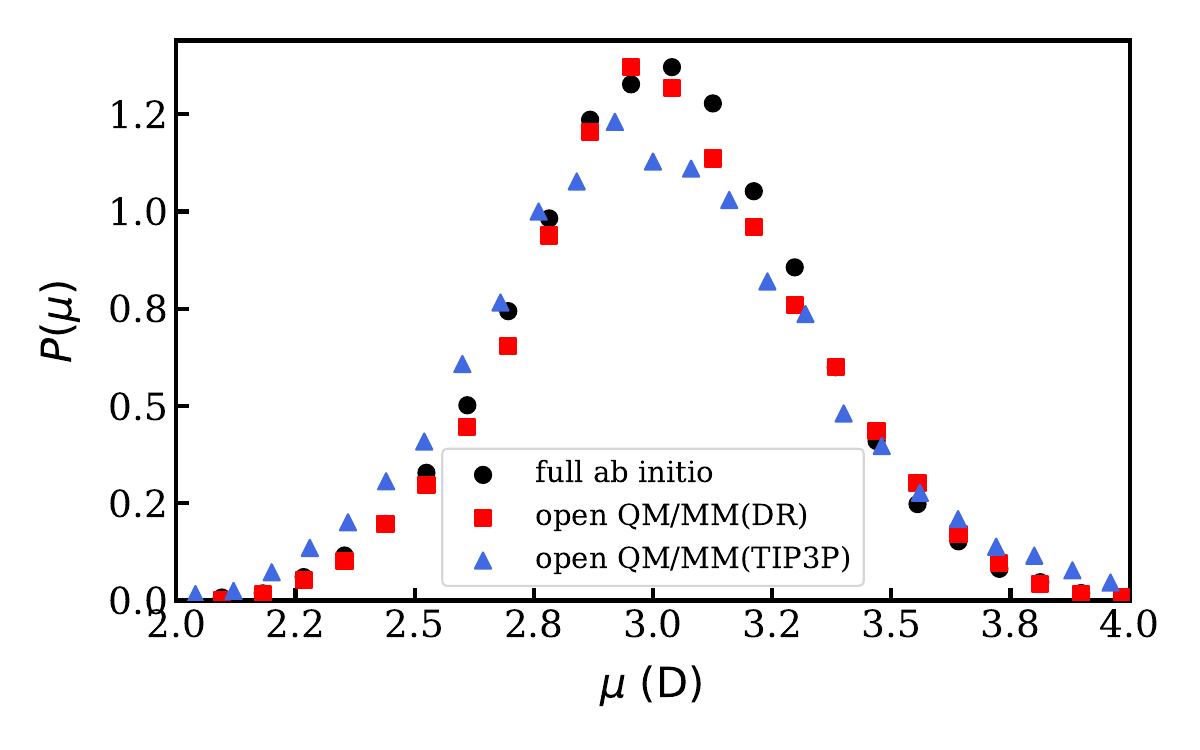}
    \caption{\justifying 
    Molecular dipole distribution $P(\mu)$ for the large QM region ($\langle N\rangle\approx440$), {comparing the full \textit{ab initio} reference, the DR reservoir, and the flexible-TIP3P reservoir. The flexible-TIP3P data, corresponding to Fig.~12 of Ref.~\cite{PanahianJand2024OpenQM}, are used with permission under the Creative Commons Attribution--NonCommercial--NoDerivatives (CC BY-NC-ND 4.0) license. Both production QM/MM simulations reproduce the reference QM molecular-dipole distribution satisfactorily.}}
    \label{fig:dip_400}
\end{figure}

The electronic-structure diagnostics reinforce this picture.The intermolecular and intramolecular decompositions of the Wannier-resolved RDFs in Fig.~\ref{fig:rdf_decomp_400} confirm that the agreement extends to both the intramolecular structure of individual water molecules and the intermolecular organization of the liquid.
 The dipole distribution in Fig.~\ref{fig:dip_400} shows that the molecular electronic response is also reproduced, as one must expect. In the large-region limit, the directly renormalized reservoir therefore acts as a physically consistent thermodynamic bath without requiring modifications to the open-boundary protocol.

\subsection{\label{190} Small QM region: $\langle N \rangle \approx 190$ molecules}

\begin{figure}[!ht]
    \centering
    \includegraphics[width=\linewidth]{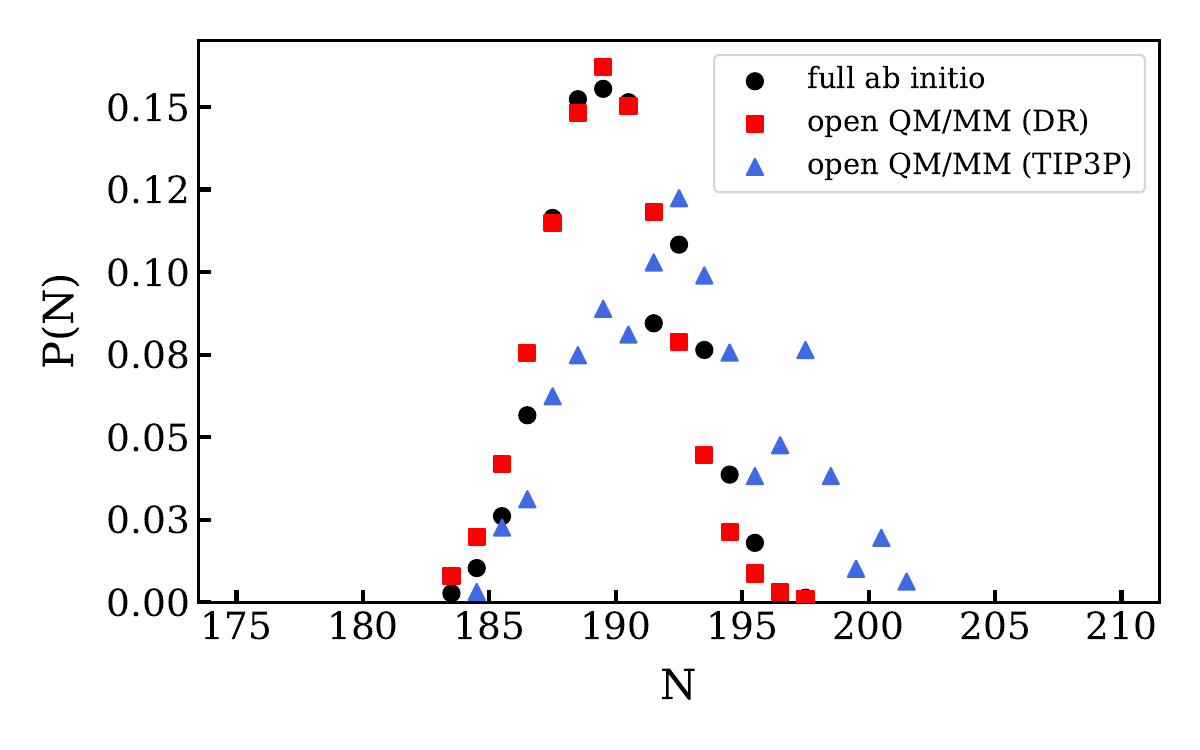}
    \caption{\justifying Particle-number distribution $P(N)$ for the 190-molecule (small) QM-region. Comparison with the full \textit{ab initio} subsystem probes whether the optimized reservoir maintains exchange of equilibrium despite the surface-to-volume ratio condition is not sufficient for a Grand Canonical ensemble. The agreement is satisfactory and thus becomes an argument supporting the (microscopic) molecular compatibility between the QM and the MM molecular description. {The flexible-TIP3P data, corresponding to Fig.~A4 of Ref.~\cite{PanahianJand2024OpenQM}, are used with permission under the Creative Commons Attribution--NonCommercial--NoDerivatives (CC BY-NC-ND 4.0) license.}}
\label{fig:pn_190}
\end{figure}

The 190-molecule case is a much stricter DR test and goes beyond the macroscopic thermodynamic role of the reservoir. In fact the larger surface-to-volume ratio makes the QM--MM interface a significant part of the environment sampled by the QM molecules. This stronger boundary sensitivity is visible in the energy-ratio fluctuation range, which increases from $\Delta R=0.048$ for the 440-molecule region to $\Delta R=0.100$ in Fig.~\ref{fig:energy_440_190}. In the previous study using the standard flexible TIP3P reservoir \cite{PanahianJand2024OpenQM}, a comparable small region exhibited clear degradation in the particle-exchange distribution and in the Wannier-resolved diagnostics. With the DR reservoir the picture improves: the particle-number distribution in Fig.~\ref{fig:pn_190} remains substantially closer to the full \textit{ab initio} subsystem, indicating better-controlled molecular exchange even when the surface-to-volume condition is no longer as favorable.

The origin of this improvement lies in the renormalization itself. Because the MM charge distribution and bonded geometry are fitted against the quantum dipole target, molecules crossing the adaptive boundary carry a classical electrostatic and structural description that is more compatible with quantum water.

\begin{figure}[!ht]
    \centering
    \begin{subfigure}[b]{0.49\linewidth}
        \centering
        \includegraphics[width=\linewidth]{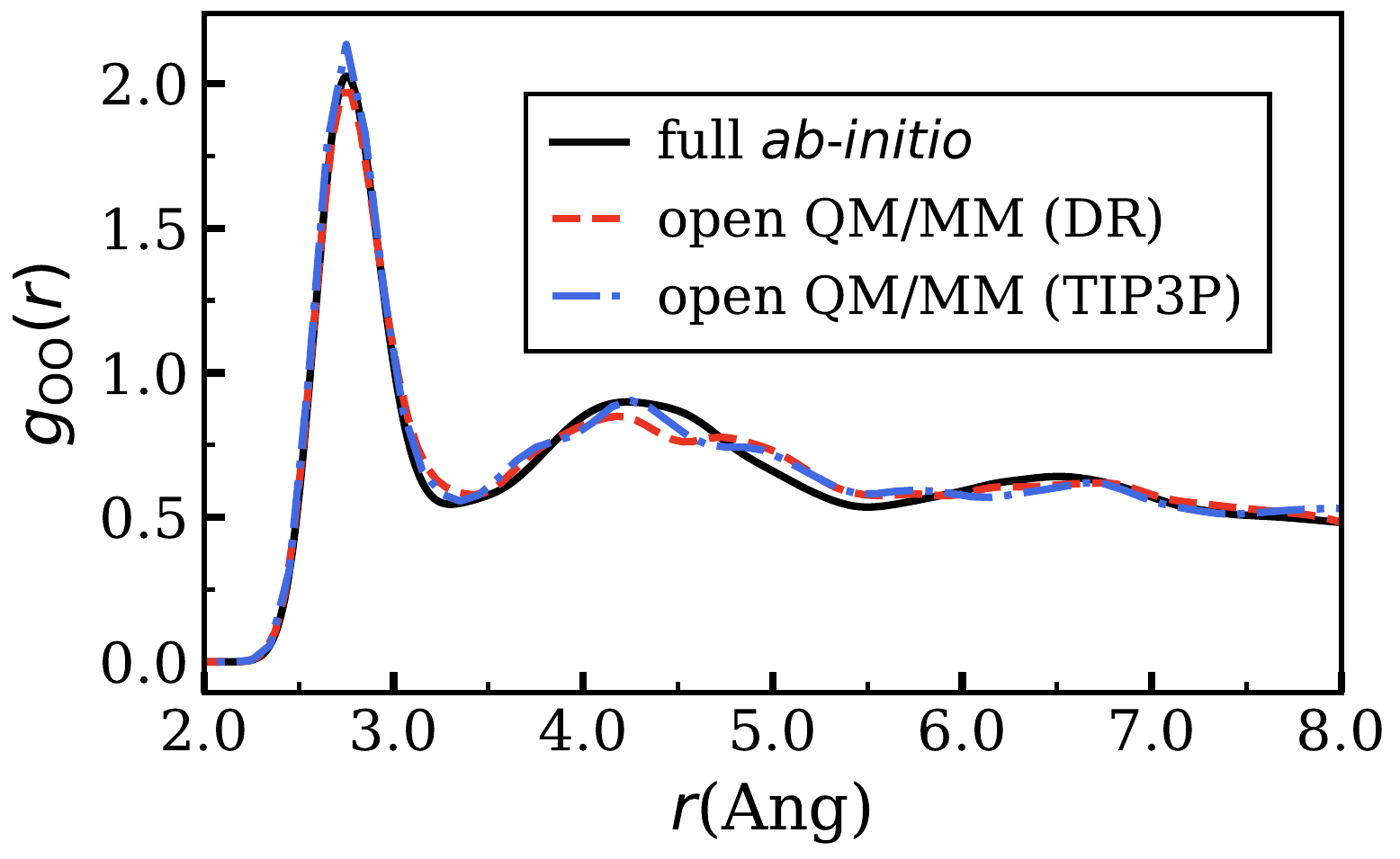}
        \caption{Oxygen--oxygen (OO) RDF}
    \end{subfigure}
    \hfill
    \begin{subfigure}[b]{0.49\linewidth}
        \centering
        \includegraphics[width=\linewidth]{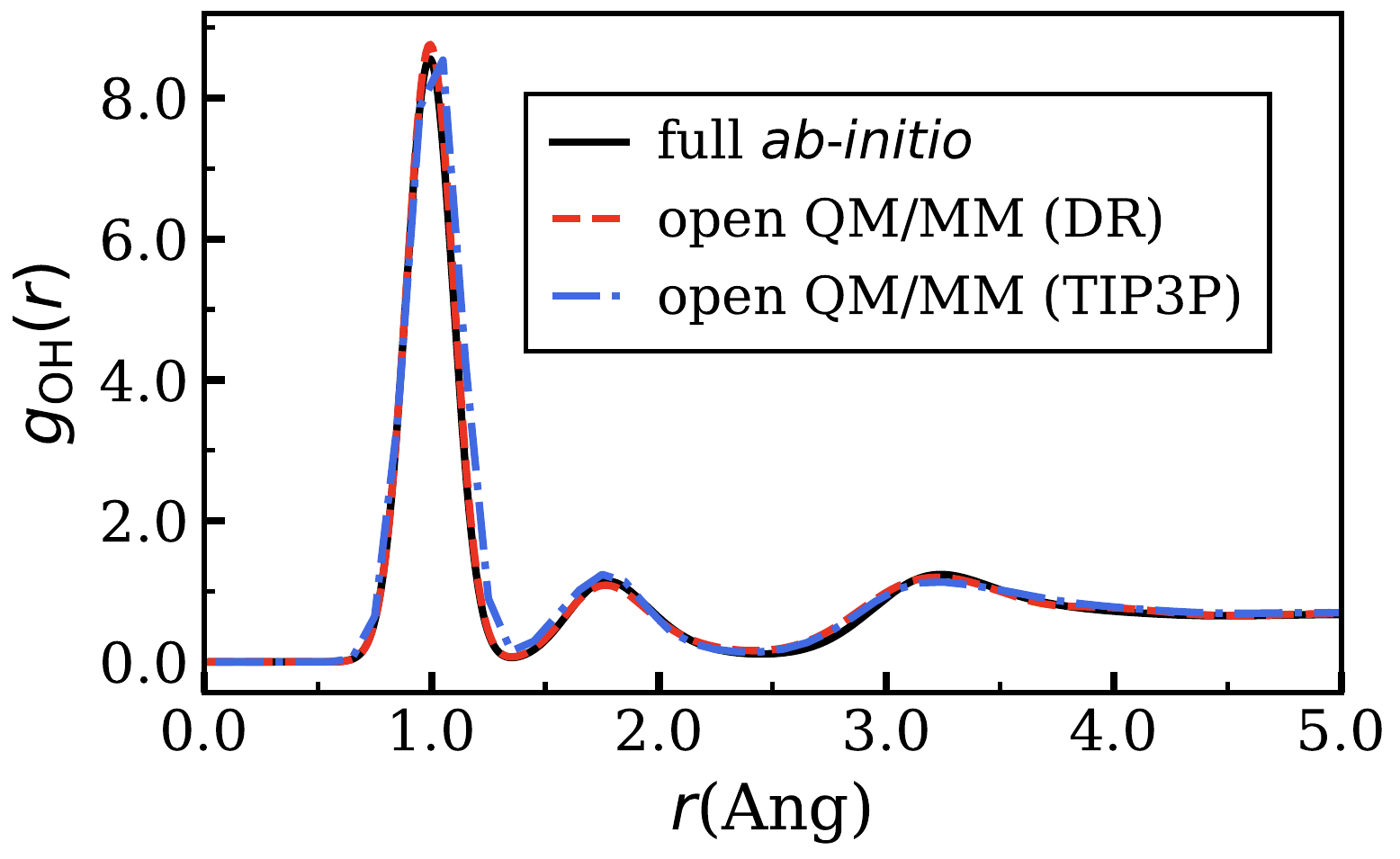}
        \caption{Oxygen--hydrogen (OH) RDF}
    \end{subfigure}
    \caption{\justifying
    Structural radial distribution functions for the 190-molecule small
    QM-region structure ($\langle N \rangle \approx 190$). These distributions
    test whether the optimized reservoir preserves the oxygen--hydrogen liquid
    structure when the QM--MM boundary effects play a more relevant role than
    in the 440-molecule case. The agreement remains satisfactory. The
    flexible-TIP3P data, corresponding to Figs.~A2 and A3 of
    Ref.~\cite{PanahianJand2024OpenQM}, are used with permission under the
    Creative Commons Attribution--NonCommercial--NoDerivatives
    (CC BY-NC-ND 4.0) license.
    }
    \label{fig:rdf_oh_190}
\end{figure}

\begin{figure}[!ht]
    \centering
    \begin{subfigure}[b]{0.47\linewidth}
        \centering
        \includegraphics[width=\linewidth]{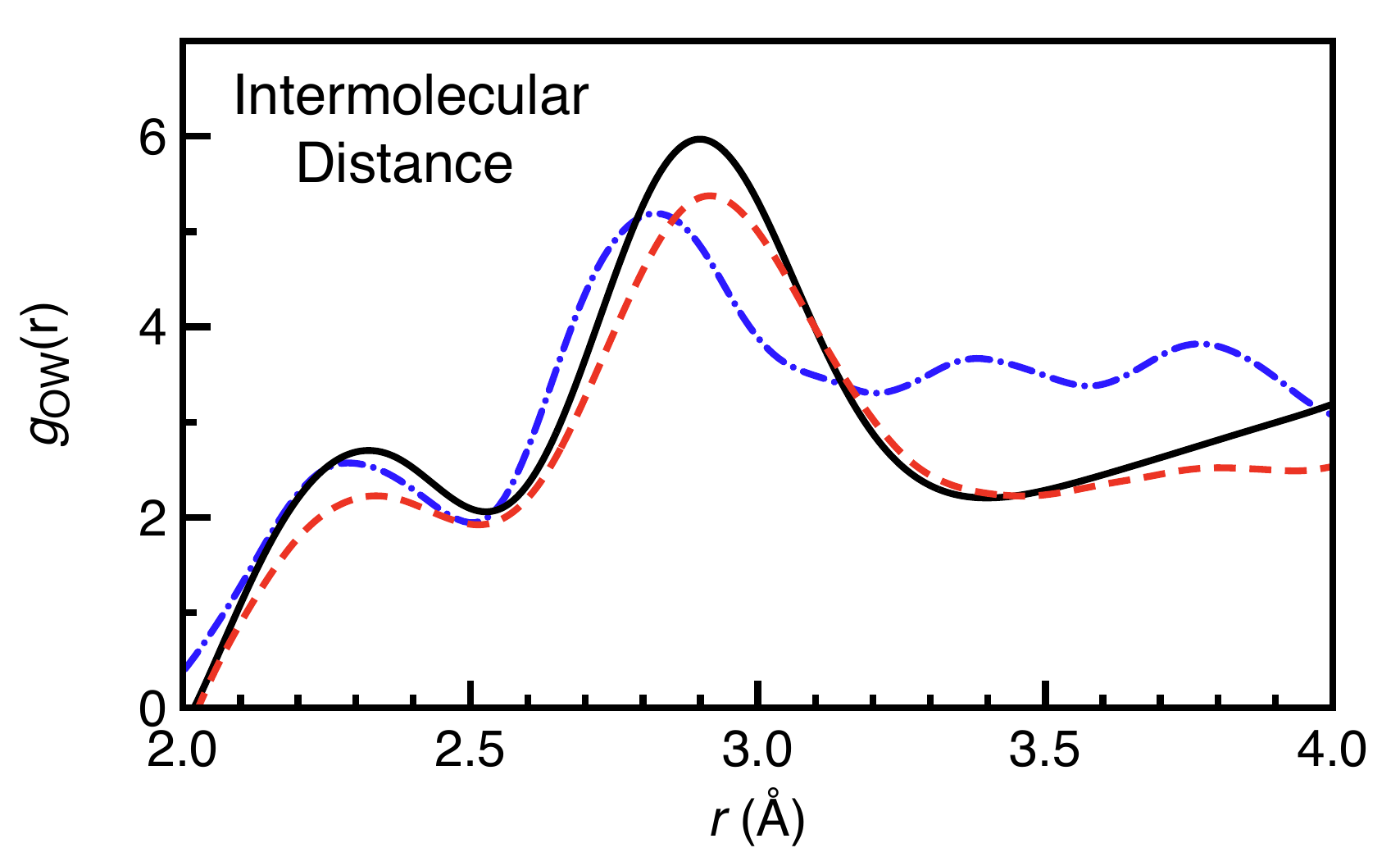}
        \caption{Intermolecular oxygen--Wannier (inter-OW)}
    \end{subfigure}
    \hfill
    \begin{subfigure}[b]{0.47\linewidth}
        \centering
        \includegraphics[width=\linewidth]{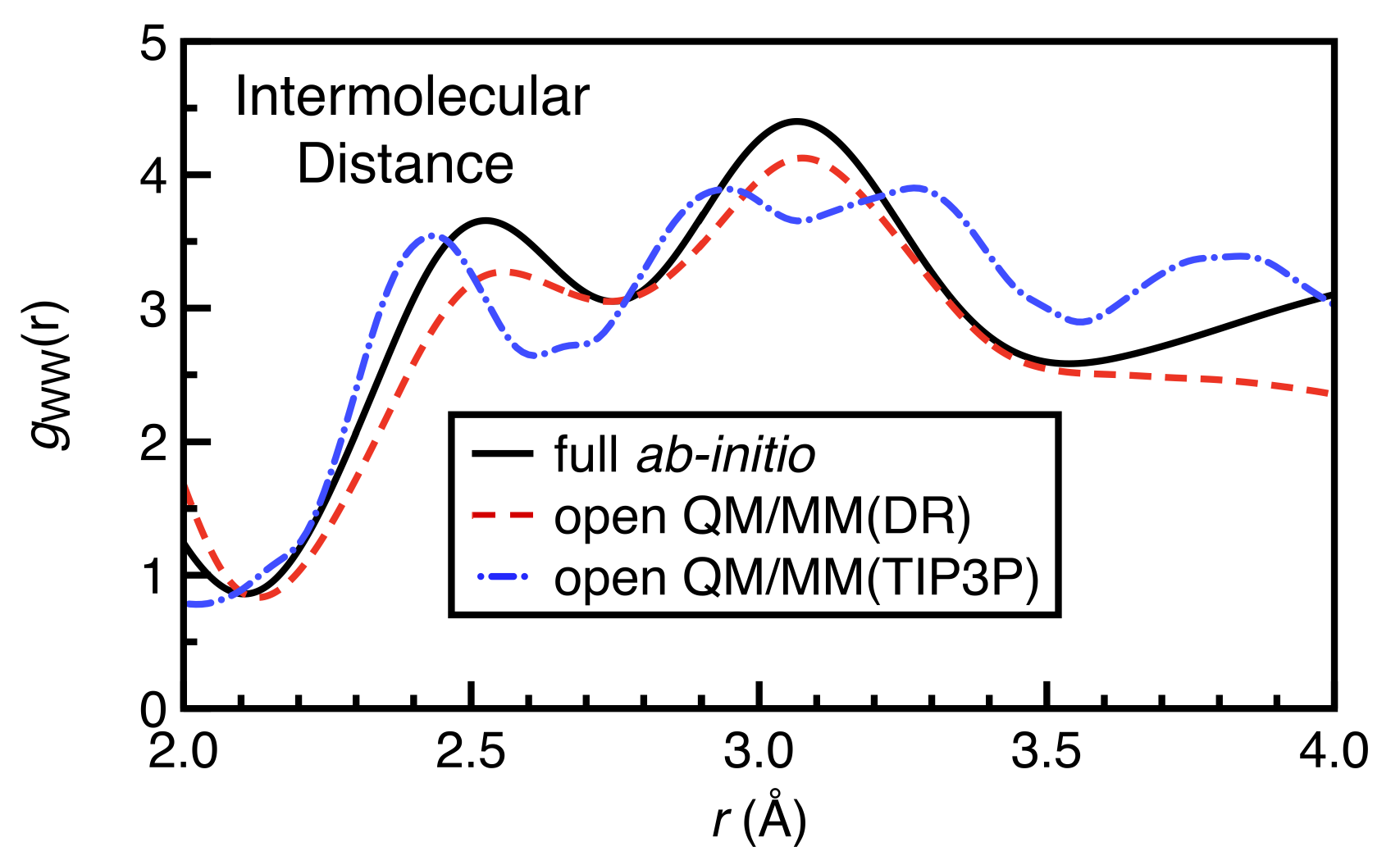}
        \caption{Intermolecular Wannier--Wannier (inter-WW)}
    \end{subfigure}
    \vspace{0.2cm}
    \begin{subfigure}[b]{0.47\linewidth}
        \centering
        \includegraphics[width=\linewidth]{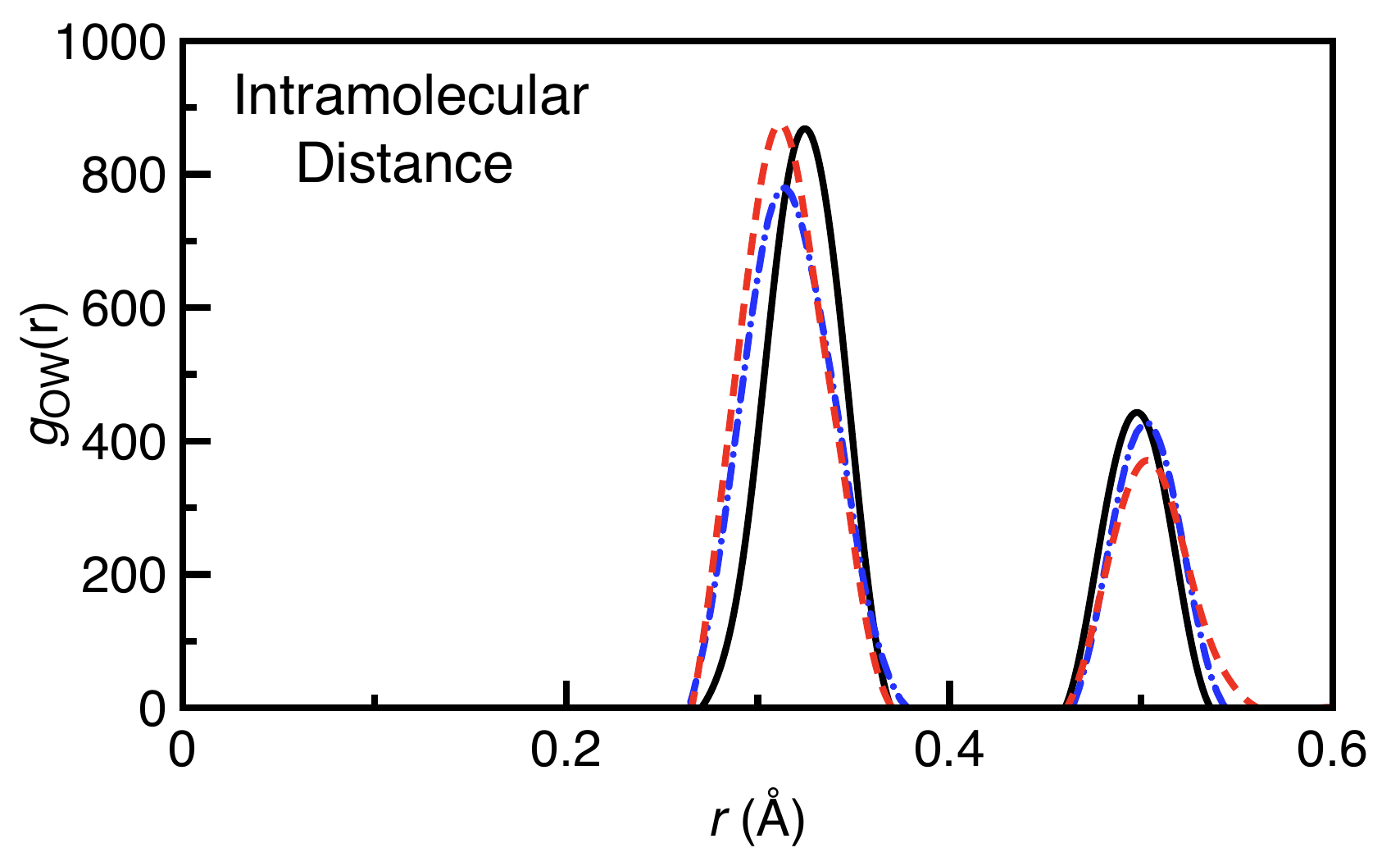}
        \caption{Intramolecular oxygen--Wannier (intra-OW)}
    \end{subfigure}
    \hfill
    \begin{subfigure}[b]{0.47\linewidth}
        \centering
        \includegraphics[width=\linewidth]{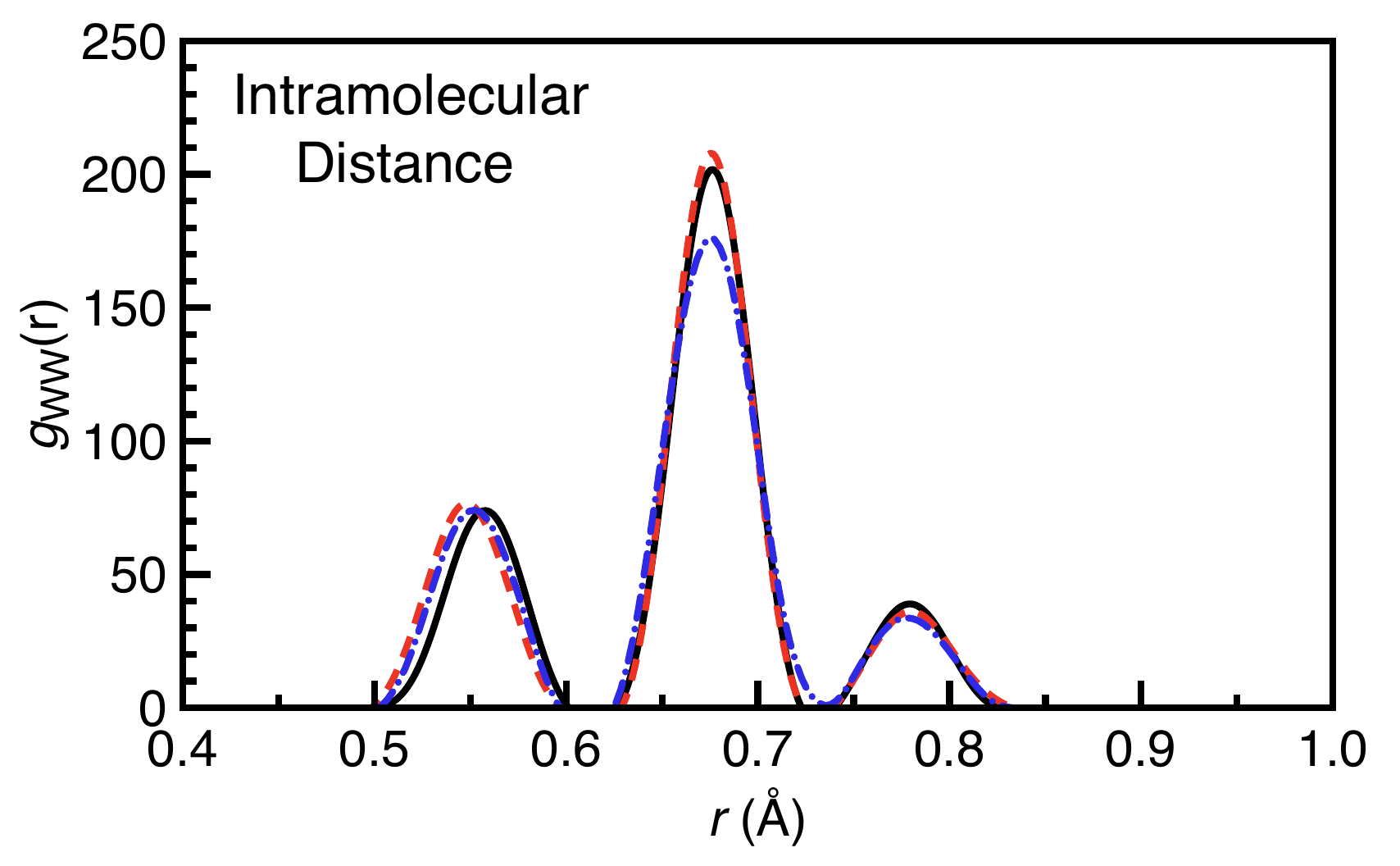}
        \caption{Intramolecular Wannier--Wannier (intra-WW)}
    \end{subfigure}
    \caption{\justifying
    Intermolecular (top) and intramolecular (bottom) decomposition of the
    Wannier-resolved RDFs for the 190-molecule small QM-region structure.
    Despite some differences, the agreement is reasonable considering that
    these distributions resolve the detailed electronic structure of individual
    molecules and of intermolecular interactions. Such fine-grained differences
    do not emerge in the molecular dipole distribution. The flexible-TIP3P data,
    corresponding to Figs.~A6, A7, A9, and A10 of
    Ref.~\cite{PanahianJand2024OpenQM}, are used with permission under the
    Creative Commons Attribution--NonCommercial--NoDerivatives
    (CC BY-NC-ND 4.0) license.
    }
    \label{fig:rdf_decomp_190}
\end{figure}

\begin{figure}[ht]
    \centering
    \includegraphics[width=\linewidth]{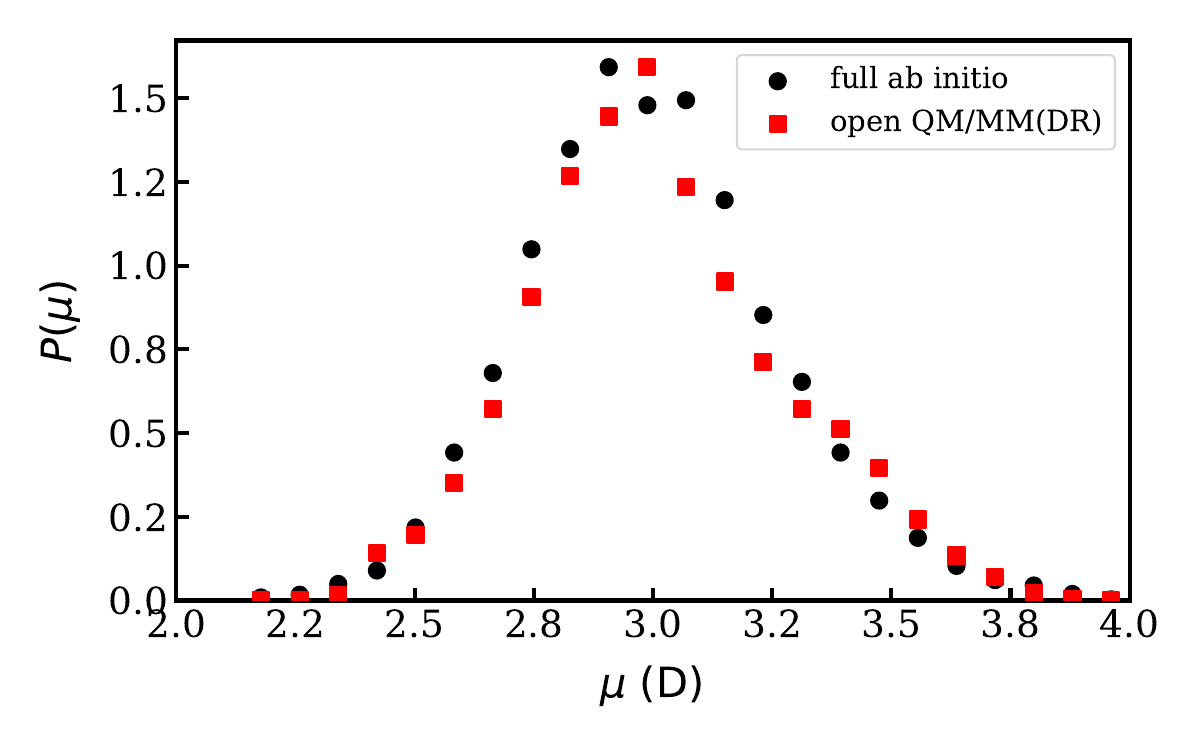}
    \caption{\justifying Molecular dipole distribution $P(\mu)$ for the $\langle N\rangle\approx190$ QM region, {comparing the DR open-QM/MM result with the corresponding full \textit{ab initio} reference.}
    }
    \label{fig:dip_190}
\end{figure}

{Figures~\ref{fig:rdf_oh_190} and~\ref{fig:rdf_decomp_190} compare the structural and Wannier-resolved RDFs obtained with the DR and flexible-TIP3P reservoirs.
Figure~\ref{fig:dip_190} shows that the molecular-dipole distribution obtained with the DR reservoir is in satisfactory agreement with the full {\it ab initio} reference. }

\section{Discussion and Conclusion}

The two main DR test cases demonstrate that the reservoir renormalization operates consistently across rather different regimes defined by the surface-to-volume ratio condition. In the large-region limit, $\langle N\rangle\approx 440$, the optimized reservoir behaves as an expected thermodynamic bath (as previously proven for less accurate molecular reservoirs \cite{PanahianJand2024OpenQM}): the small energy-ratio fluctuation and the agreement of $P(N)$, OO/OH structural RDFs, Wannier-resolved RDFs, and dipole distribution with the \textit{\it ab initio} subsystem confirm a full Grand Canonical consistency. In the smaller-region test, $\langle N\rangle\approx 190$, the interface effects become statistically significant and one is, strictly speaking, no more in a full Grand Canonical regime. Yet the same DR reservoir shows a satisfactory level of  molecular consistency when running  the same diagnostics (contrary to the case of Ref. \citenum{PanahianJand2024OpenQM}).
The two aspects discussed above are not equivalent: the first discussion about the large QM region concerns essentially a macroscopic thermodynamic check, whereas the second aspects, that is the smaller QM region, concerns more specifically the level of molecular compatibility of the reservoir w.r.t. quantum molecular water. This is due to the fact that molecules continuously cross the QM-MM boundary and the boundary effects play a major role in the properties of the QM region in its smaller size. In fact, if the small QM system behaves as a small subsystem of a large {\it ab initio} system, the natural implication is that the MM molecules that constitute the environment must carry microscopic characteristics similar to the quantum ones.

The contrast with the previous flexible TIP3P study \cite{PanahianJand2024OpenQM} makes the effect of renormalization concrete. In that work, for a QM region of 440-molecules one could already have a satisfactory Grand Canonical statistics with the non optimized TIP3P reservoir (i. e. a satisfactory macroscopic thermodynamic bath). However for the smaller QM region containing about 192 molecules one could observe a clearly different particle-exchange distribution and far more pronounced deviations in the Wannier-resolved diagnostics. The 190-molecule DR results reported here improve, up to a {reasonable} level,  {the particle-number fluctuations and the Wannier intramolecular radial distribution functions, compared to the results of Ref.~\cite{PanahianJand2024OpenQM}, at the same time they provide a satisfactory agreement for molecular dipole distribution compared to the \textit{ab initio} reference, suggesting a genuine improvement of the quality of the  reservoir.}

{The physical origin of the improvement can be understood from the role of the molecular dipole as the leading anisotropic electrostatic descriptor of neutral water. In the flexible TIP3P reservoir, molecules crossing the adaptive QM/MM boundary carry a generic classical charge distribution calibrated primarily for bulk thermodynamic properties. In an adaptive QM/MM simulation, molecules repeatedly cross this boundary and change their representation. If the classical and quantum molecular dipoles are substantially different, this transition changes the electrostatic interactions and the effective free-energy cost of molecular exchange. Reducing this mismatch therefore improves the insertion and removal balance at the interface and can lead to a more consistent particle number distribution.
\\
Dipolar interactions also influence molecular orientation,
hydrogen-bond organization, and the local electric field acting on the QM molecules. A more compatible reservoir can consequently reduce distortions in the structural RDFs and in the electronic polarization represented by the Wannier resolved RDFs. Nevertheless, matching the dipole distribution does not guarantee agreement in these observables; rather, it removes an important low order electrostatic mismatch. The particle-number, structural, and Wannier resolved distributions
therefore provide independent tests of reservoir compatibility.}

In the Appendix~\ref{app:qm50_results} we report results for an additional $\langle N\rangle\approx 50$. It obviously probes an extreme regime: at this size, the boundary interaction is comparable to the internal contribution of the QM subsystem. The surface-to-volume ratio condition required for a rigorous Grand Canonical description is therefore strongly violated.

In this case, the calculation tests whether the optimized reservoir remains locally (highly) compatible with the quantum description when nearly every QM molecule is influenced by the adaptive interface.

The additional 50-molecule calculation shows that meaningful local structural and electronic features can remain reasonably preserved even in the extreme regime in which interface effects play a major role. At the same time, the larger deviations in the particle-number, Wannier-resolved, and molecular-dipole distributions identify the limitations of a dipole-renormalized three-site description in this extreme boundary-dominated regime. The calculation therefore defines a practical molecular-compatibility limit, w.r.t. the QM molecular description, of the present molecular model of the reservoir based on the dipole distribution only.

However, it must be clarified that simulations with regions smaller than the
minimal size required by the Grand Canonical criterion can only be considered qualitative studies of local properties. Averages of physical observables over such small regions do not have a rigorous statistical-mechanical meaning, in contrast to simulations with QM regions large enough to satisfy
the Grand Canonical criterion, i.e. (strictly speaking) $\langle N\rangle\ge 440$.

{The present dipole-based strategy is not expected to be sufficient for all molecular systems. For ionic liquids, electrolytes, organic solvents, and biomolecular environments, higher multipoles, polarizabilities, charge transfer, or other system-specific quantum descriptors may be required. The general feature of the framework is therefore the incorporation of relevant quantum-derived information into the reservoir, rather than the universal use of the molecular dipole alone. A general aspect of the proposed strategy, underlined before, is that the method itself can be used for identifying the minimal level of quantum-derived information needed in building a classical model of reservoir compatible with the QM region; this is not only a technical information for running efficient simulations but it actually is a straightforward way to rationalize the physics of a classical molecular model.}

Nevertheless, for local properties such as bond rearrangement, dissociation paths, or reaction pathways, such simulations, e.g. $\langle N\rangle\approx 190$ or even {{$\langle N\rangle\approx 50$}}, can be useful and efficient. They reduce the computational costs and can help to identify physical trends which can be  subsequently verified by more expensive calculations using larger QM regions that satisfy the appropriate statistical-mechanical conditions. In conclusion, we have employed the theoretical protocol to define physically consistent open systems that exchange particles with a reservoir. We have applied it to the case of the open QM-MM computational approach and propose to use the size of the QM region as a probe of molecular consistency between the QM and classical molecular representation. Our results show that one can indeed use such a procedure to systematically build molecular models which carry relevant aspects of the quantum molecular nature. In the current era of machine learning in molecular simulation \cite{cecilia} our proposal can be taken as a theoretical tool to possibly build classical models with, as a matter of fact, arbitrary molecular compatibility with the quantum molecules. The advantage in this case is that one can use our approach and the concept of compatibility to identify those molecular aspects that are dominant in the characterization of the "quantumness" of a molecule by selecting the information to feed into the learning process. 


\appendix

\section{Convergence of the dipole-renormalized reservoir}
\label{app:optimization_convergence}

The construction of the dipole-renormalized reservoir is described in Sec.~\ref{application}. Here we report the convergence behavior of the optimization and the retained force-field parameters used in the production open-QM/MM simulations.
{The complete training procedure involved approximately 800 parameter-update steps and required about 1152 allocated core-hours. This represents a one-time parameterization cost. For a 30~ps full-MM benchmark of 2960 water molecules, the DR and flexible TIP3P simulations required 21.86 and 22.42~CPU core-hours, respectively, using one node with 24 MPI tasks and four CPU cores per task. Because both models have the same analytical form, their production costs are essentially equivalent.}

\begin{figure}[!ht]
    \centering
    \includegraphics[width=0.72\linewidth]{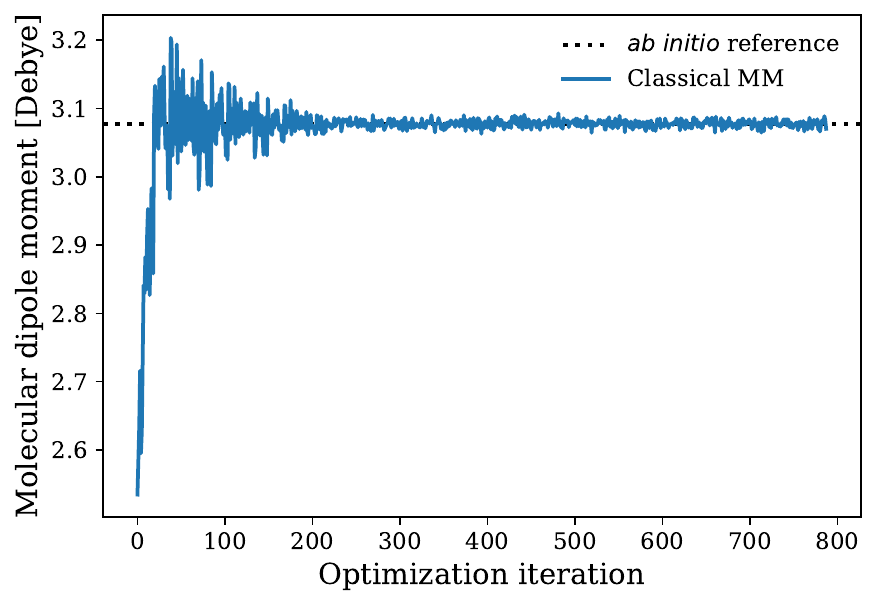}
    \caption{ \justifying
    Convergence of the molecular dipole during the stochastic optimization of the dipole-renormalized reservoir. The dotted horizontal line indicates the mean molecular dipole of the full \textit{ab initio} reference trajectory, while the solid line shows the corresponding molecular dipole
    obtained from the candidate classical reservoir during optimization. 
    {Intermediate values correspond to trial MM parameter sets; only the final converged {set of parameters} was used in the production QM/MM simulations.}
    }
    \label{fig:app_dipole_convergence}
\end{figure}

Figure~\ref{fig:app_dipole_convergence} shows the convergence of the molecular dipole during the stochastic optimization. The molecular dipole of the candidate classical reservoir approaches the mean value of the full \textit{ab initio} reference trajectory and remains stable after the initial
optimization stage. This confirms that the retained Amber-compatible {set of parameters} reproduces the target low-order electronic descriptor used in the renormalization procedure.

\begin{figure}[!htbp]
    \centering
    \includegraphics[width=0.85\linewidth]{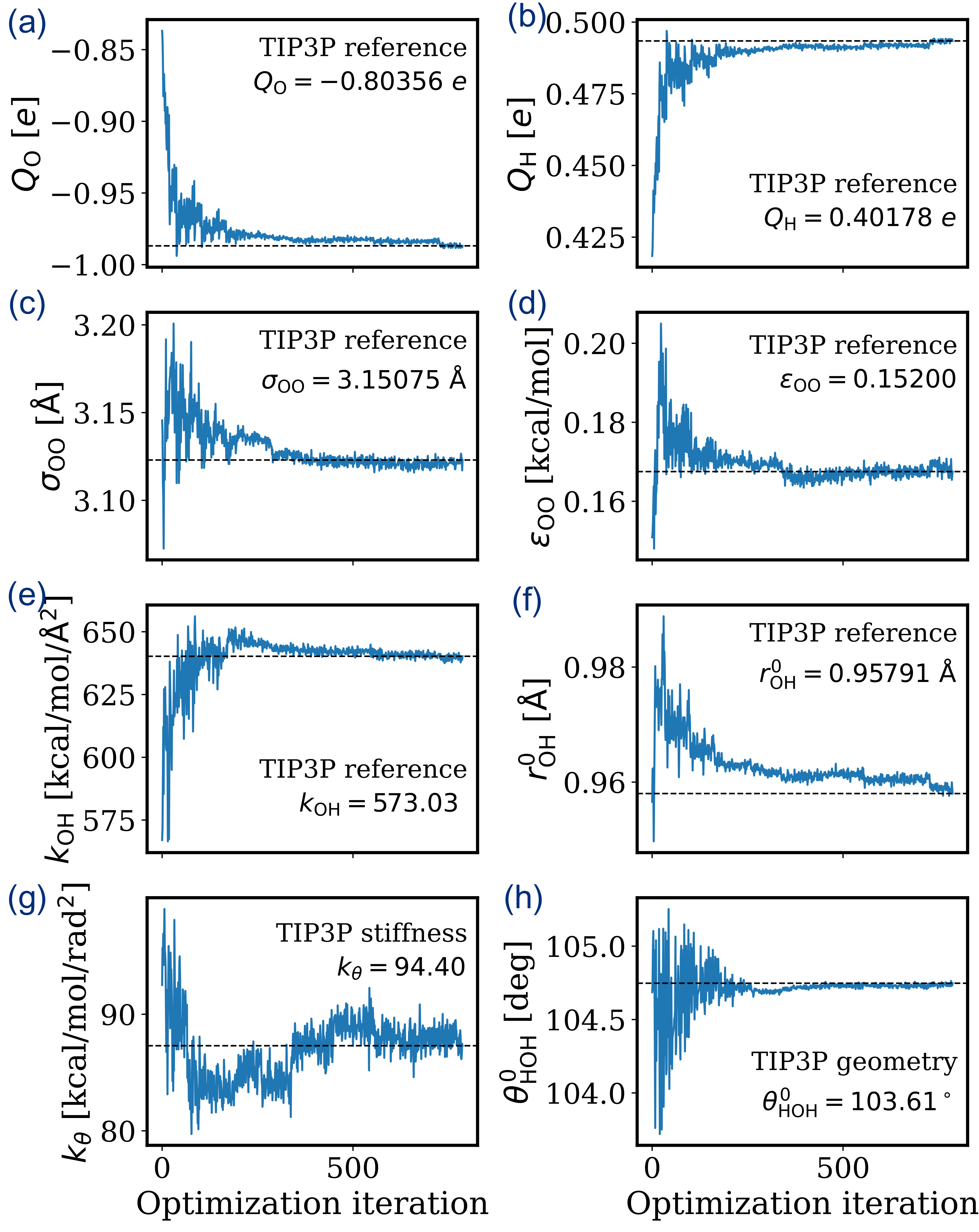}
    \caption{\justifying
    Optimization trajectories of the force-field parameters of the
    dipole-renormalized reservoir: (a) oxygen partial charge $Q_{\mathrm O}$;
    (b) hydrogen partial charge $Q_{\mathrm H}$;
    (c) oxygen--oxygen Lennard--Jones size parameter $\sigma_{\mathrm{OO}}$;
    (d) oxygen--oxygen Lennard--Jones energy parameter $\epsilon_{\mathrm{OO}}$;
    (e) O--H bond force constant $k_{\mathrm{OH}}$;
    (f) equilibrium O--H bond length $r_{\mathrm{OH}}^0$;
    (g) H--O--H angular force constant $k_\theta$; and
    (h) equilibrium H--O--H angle $\theta_{\mathrm{HOH}}^0$.
    The dashed horizontal line in each panel indicates the final retained
    value used in the production open-QM/MM simulations.
    }
    \label{fig:app_parameter_convergence}
\end{figure}

Figure~\ref{fig:app_parameter_convergence} reports the corresponding optimization trajectories of the force-field parameters. After an initial exploratory regime, the oxygen and hydrogen partial charges, Lennard-Jones parameters, bonded force constants, equilibrium bond distance, and equilibrium angle approach stable plateaus. The final retained values were taken from this converged region and used to construct the MM reservoir in the open-QM/MM production simulations.

The retained parameter values are indicated by the dashed horizontal lines in Fig.~\ref{fig:app_parameter_convergence}. Hydrogen Lennard-Jones parameters were kept equal to zero, and the hydrogen partial charge follows from charge neutrality, $Q_{\mathrm H}=-Q_{\mathrm O}/2$.

The convergence of both the target molecular dipole and the underlying force-field parameters shows that the final {set of parameters} was selected from a stable optimization regime.

\section{Extreme small-region compatibility test:
$\langle N\rangle\approx50$}
\label{app:qm50_results}

In addition to the two principal QM-region sizes discussed in the main text, we considered an extreme small region containing on average approximately 50 water molecules. The approximate radius of this region is $7.0~\text{\AA}$.

As discussed in Sec.~\ref{results}, this system strongly violates the surface-to-volume condition required for a rigorous (or within certain precision, approximate) Grand Canonical interpretation. It can be therefore used only to assess whether the optimized reservoir offers a certain level of microscopic compatibility at molecular and electronic level so that one may trust the study of very localized phenomena done with this open QM-MM set up.

\begin{figure}[!ht]
    \centering
    \includegraphics[width=0.75\linewidth]{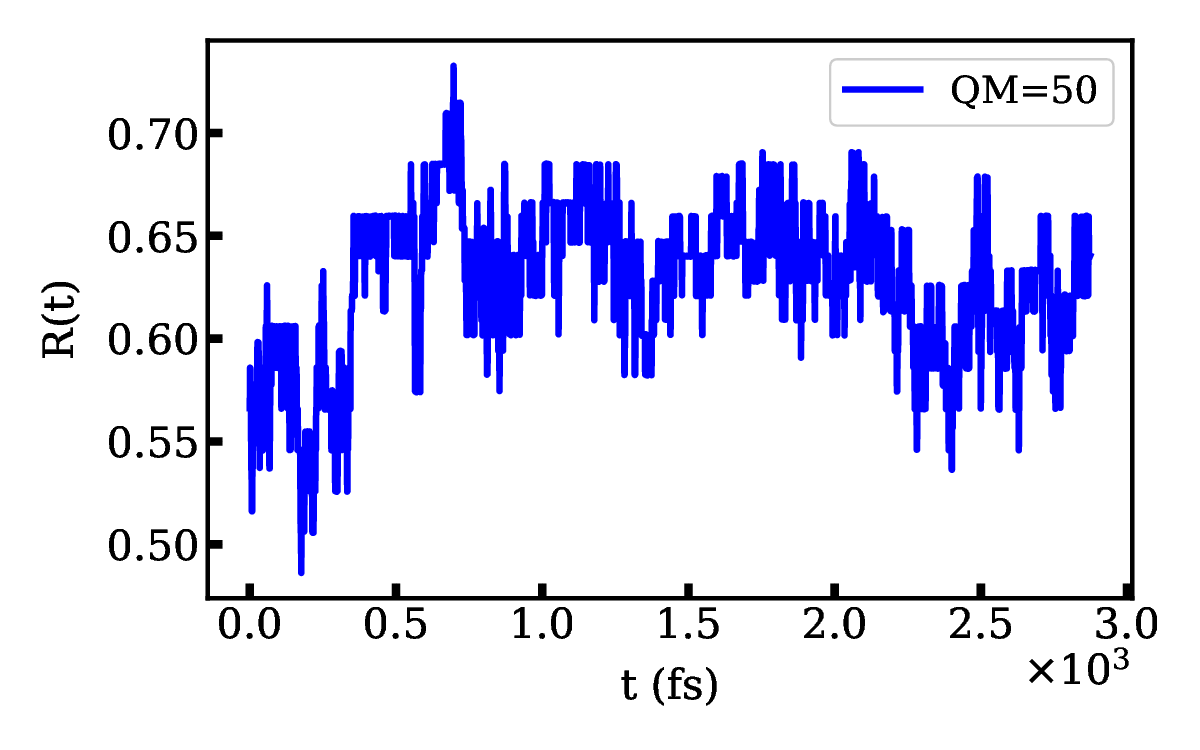}
    \caption{\justifying Time evolution of the energy ratio $R(t)$ for the extreme small QM-region structure ($\langle N\rangle\approx50$). The ratio remains far from unity and shows stronger fluctuations than in the $\langle N\rangle\approx440$ and $\langle N\rangle\approx190$ cases, confirming that the boundary interaction is no longer negligible compared with the internal QM contribution. This system is therefore interpreted as a local molecular-compatibility test rather than as a Grand Canonical open-system validation.}
    \label{fig:app_energy_50}
\end{figure}

The energy-ratio diagnostic provides a direct indication of the boundary-dominated character of this small subsystem. As shown in Fig.~\ref{fig:app_energy_50}, $R(t)$ remains significantly below unity and fluctuates over a wider range than in the larger QM regions. The corresponding fluctuation range is $ \Delta R \approx 0.2695,$ which is substantially larger than the values obtained for the $\langle N\rangle\approx440$ and $\langle N\rangle\approx190$ regions. This confirms that the 50-molecule region is the most strongly interface-dominated case considered in this work.
\begin{figure}[!ht]
    \centering

    \begin{subfigure}[b]{0.48\linewidth}
        \centering
        \includegraphics[width=\linewidth]{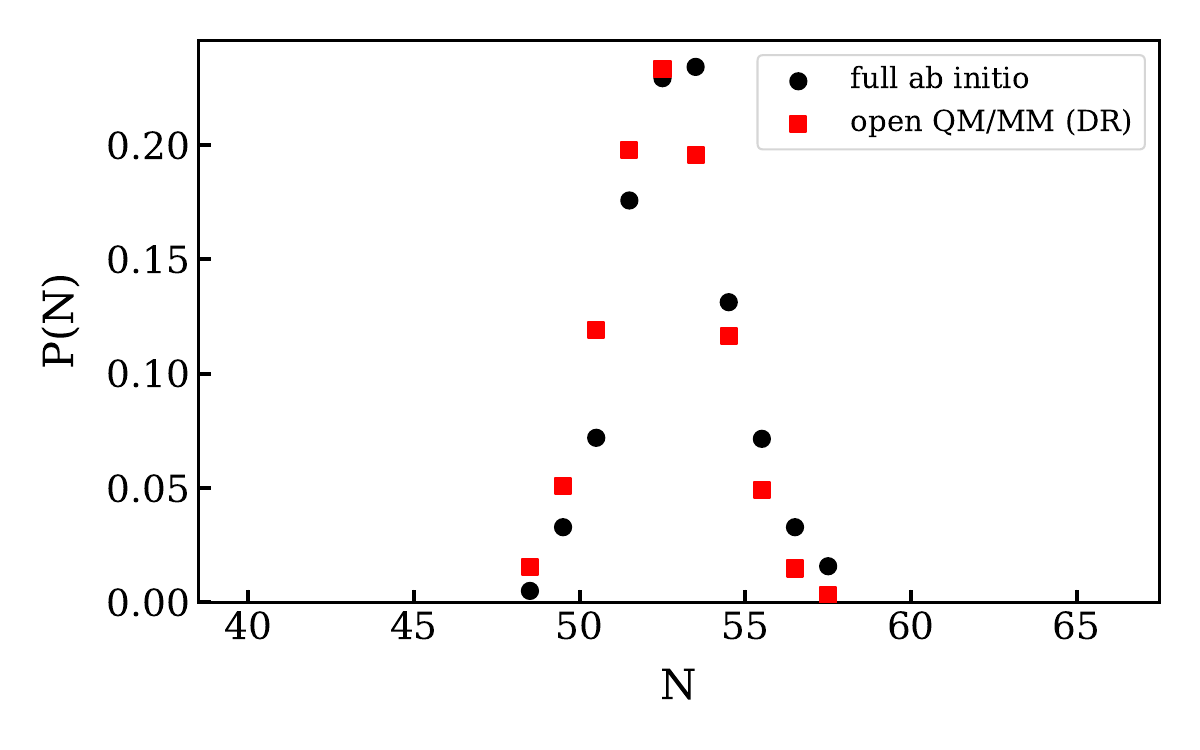}
        \caption{Particle-number distribution}
        \label{fig:app_pn_50}
    \end{subfigure}
    \hfill
    \begin{subfigure}[b]{0.48\linewidth}
        \centering
        \includegraphics[width=\linewidth]{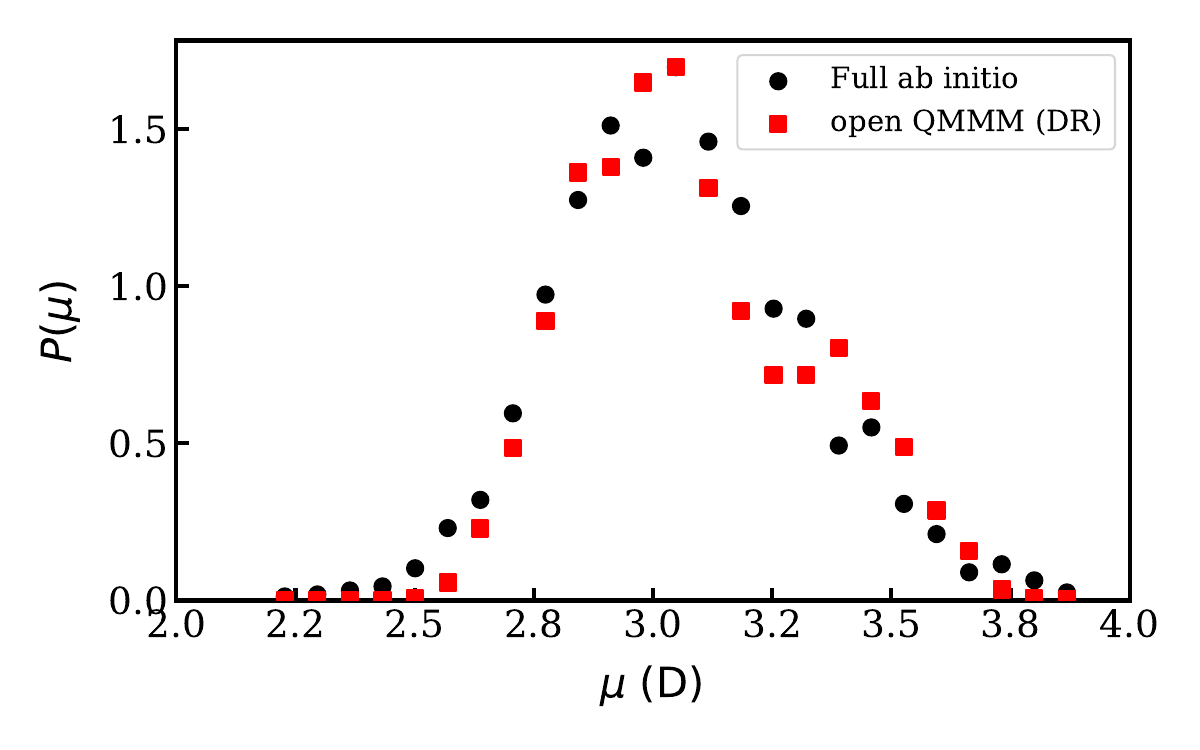}
        \caption{Molecular dipole distribution}
        \label{fig:app_dipole_50}
    \end{subfigure}

    \caption{ \justifying
    Particle-number and molecular-dipole distributions for the $\langle N\rangle\approx50$ QM region.
    Panel (a) shows $P(N)$, used here as a diagnostic of the stability of molecular exchange rather than as evidence of Grand Canonical statistics. Panel (b) shows the molecular dipole distribution $P(\mu)$, which probes the local electronic polarization of the QM water molecules.
    }
    \label{fig:app_pn_dipole_50}
\end{figure}

The particle-number distribution in Fig.~\ref{fig:app_pn_dipole_50}(a) remains centered near the target region size. This shows that the adaptive partition remains dynamically stable and that the small QM region is neither depleted nor subject to uncontrolled density accumulation. 

The molecular dipole distribution in Fig.~\ref{fig:app_pn_dipole_50}(b) retains the characteristic dipole scale of liquid water. The remaining broadening and differences in the distribution show, however, that a dipole-matched three-site reservoir cannot fully represent all higher-order polarization and boundary-specific effects in this extreme small-region regime.

\begin{figure}[!ht]
    \centering

    \begin{subfigure}[b]{0.48\linewidth}
        \centering
        \includegraphics[width=\linewidth]{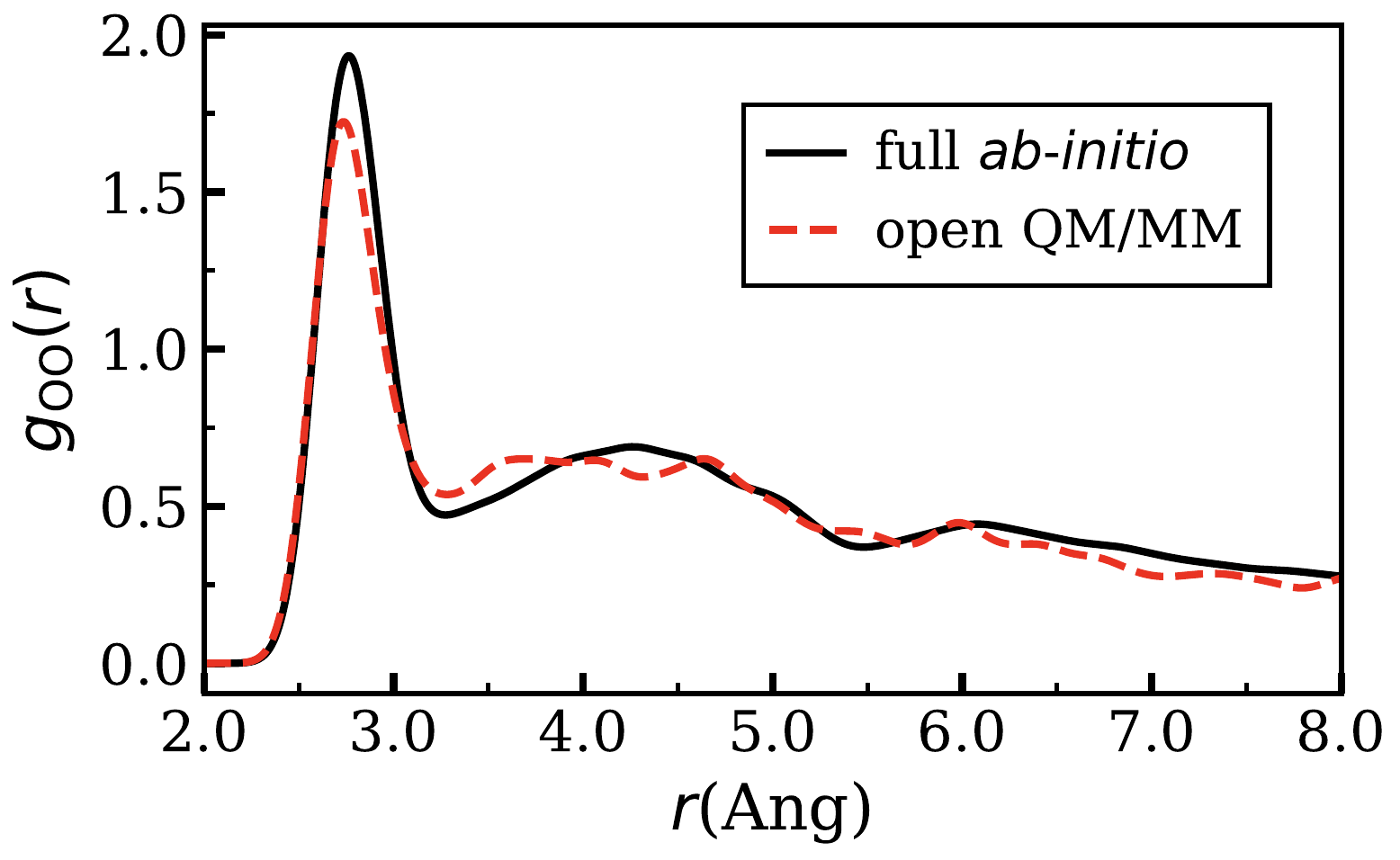}
        \caption{Oxygen--oxygen RDF}
        \label{fig:app_rdf_oo_50}
    \end{subfigure}
    \hfill
    \begin{subfigure}[b]{0.48\linewidth}
        \centering
        \includegraphics[width=\linewidth]{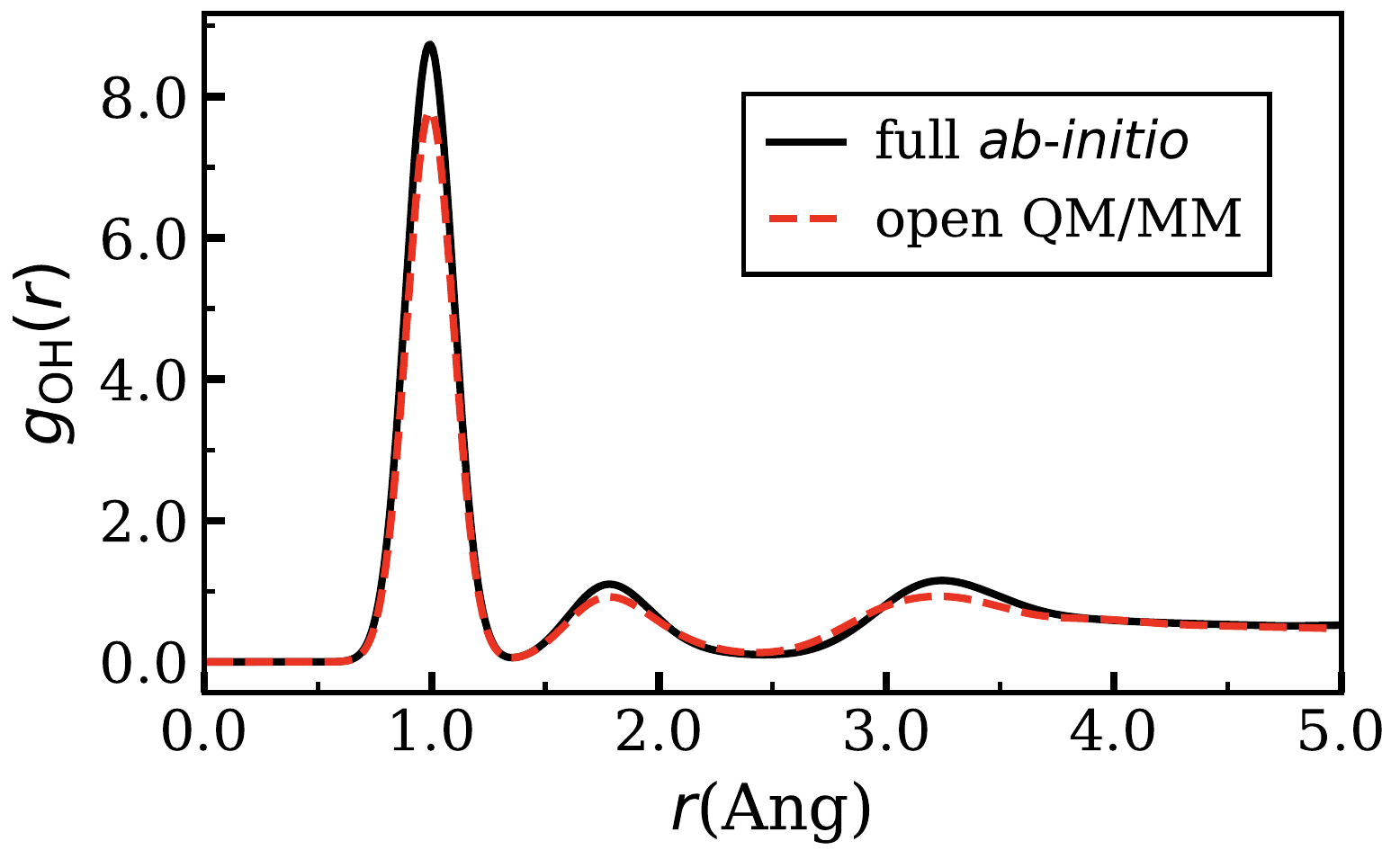}
        \caption{Oxygen--hydrogen RDF}
        \label{fig:app_rdf_oh_50}
    \end{subfigure}

    \caption{ \justifying
    Structural radial distribution functions for the $\langle N\rangle\approx50$ QM region. Panels (a) and (b) show the oxygen--oxygen and oxygen--hydrogen distributions, respectively.
    }
    \label{fig:app_structural_rdfs_50}
\end{figure}

The structural RDFs in Fig.~\ref{fig:app_structural_rdfs_50} retain the main features of liquid water. In particular, the principal OO and OH peaks remain in the same distance ranges as the corresponding \textit{ab initio} subsystem. The residual differences in peak shape and longer-range structure are larger than in the principal QM-region calculations, consistent with the stronger influence of the QM--MM interface.





\begin{figure}[!ht]
    \centering

    \begin{subfigure}[b]{0.47\linewidth}
        \centering
        \includegraphics[width=\linewidth]{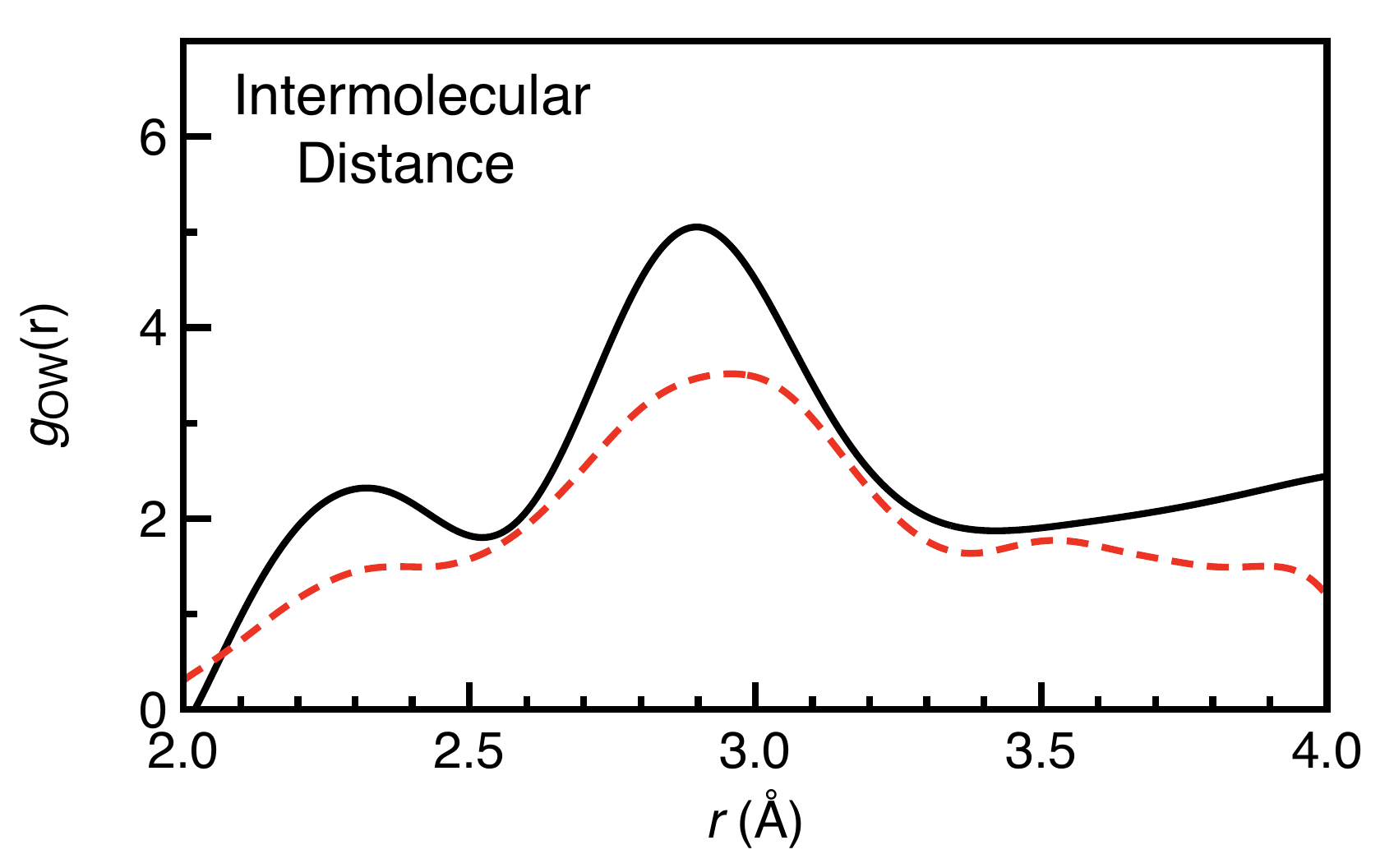}
        \caption{Intermolecular oxygen--Wannier}
        \label{fig:app_inter_xo_50}
    \end{subfigure}
    \hfill
    \begin{subfigure}[b]{0.47\linewidth}
        \centering
        \includegraphics[width=\linewidth]{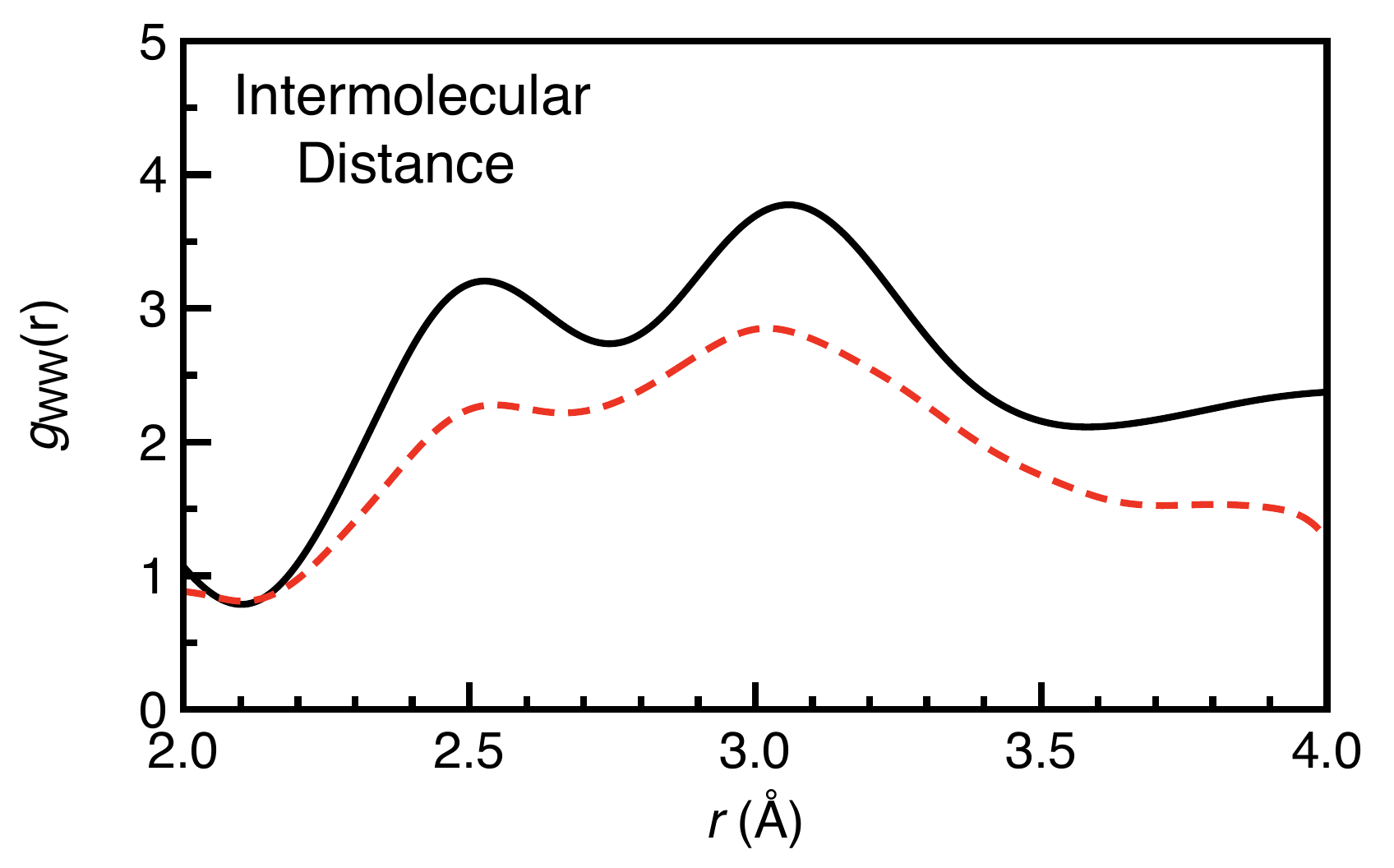}
        \caption{Intermolecular Wannier--Wannier}
        \label{fig:app_inter_xx_50}
    \end{subfigure}

    \vspace{0.2cm}

    \begin{subfigure}[b]{0.47\linewidth}
        \centering
        \includegraphics[width=\linewidth]{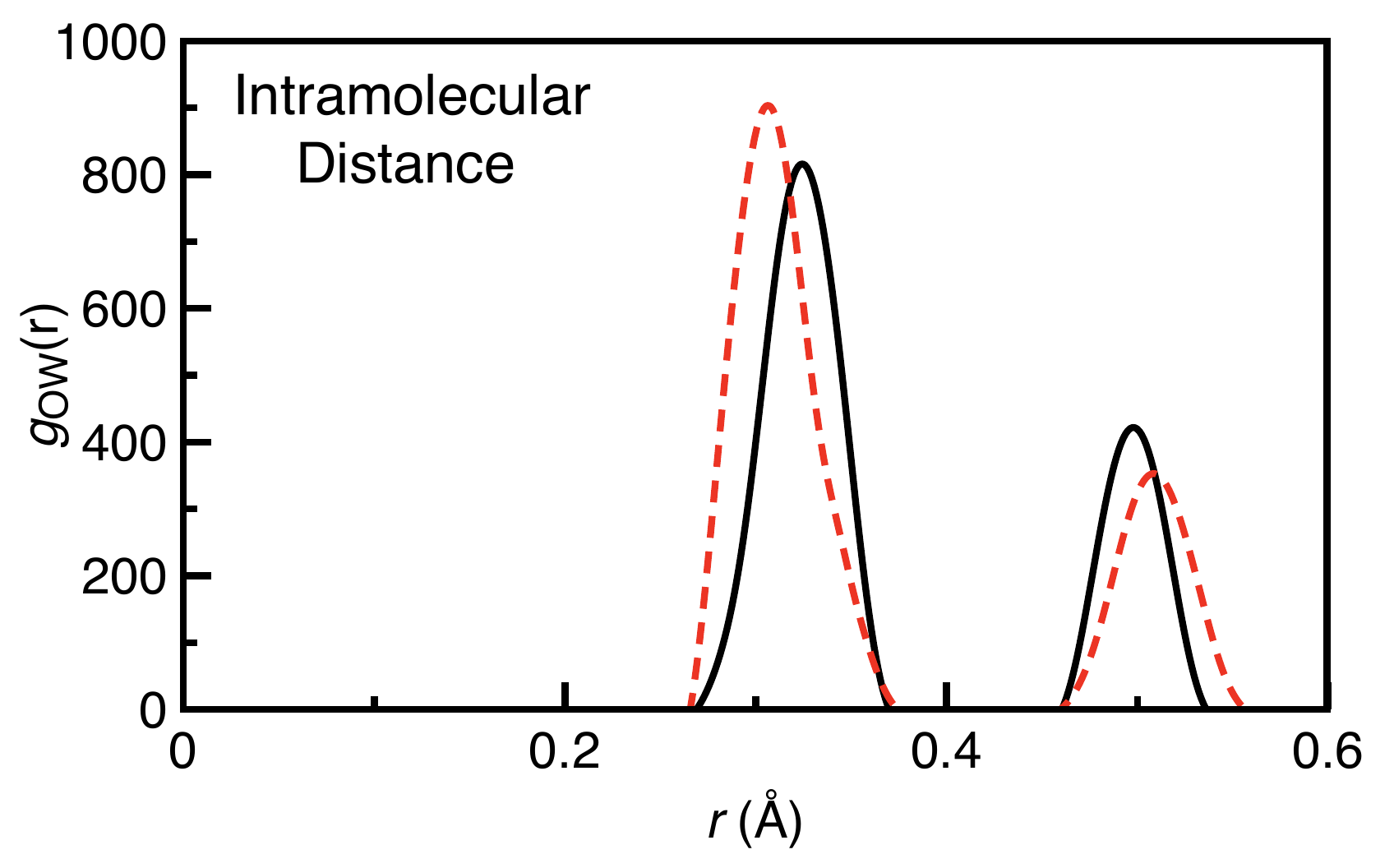}
        \caption{Intramolecular oxygen--Wannier}
        \label{fig:app_intra_xo_50}
    \end{subfigure}
    \hfill
    \begin{subfigure}[b]{0.47\linewidth}
        \centering
        \includegraphics[width=\linewidth]{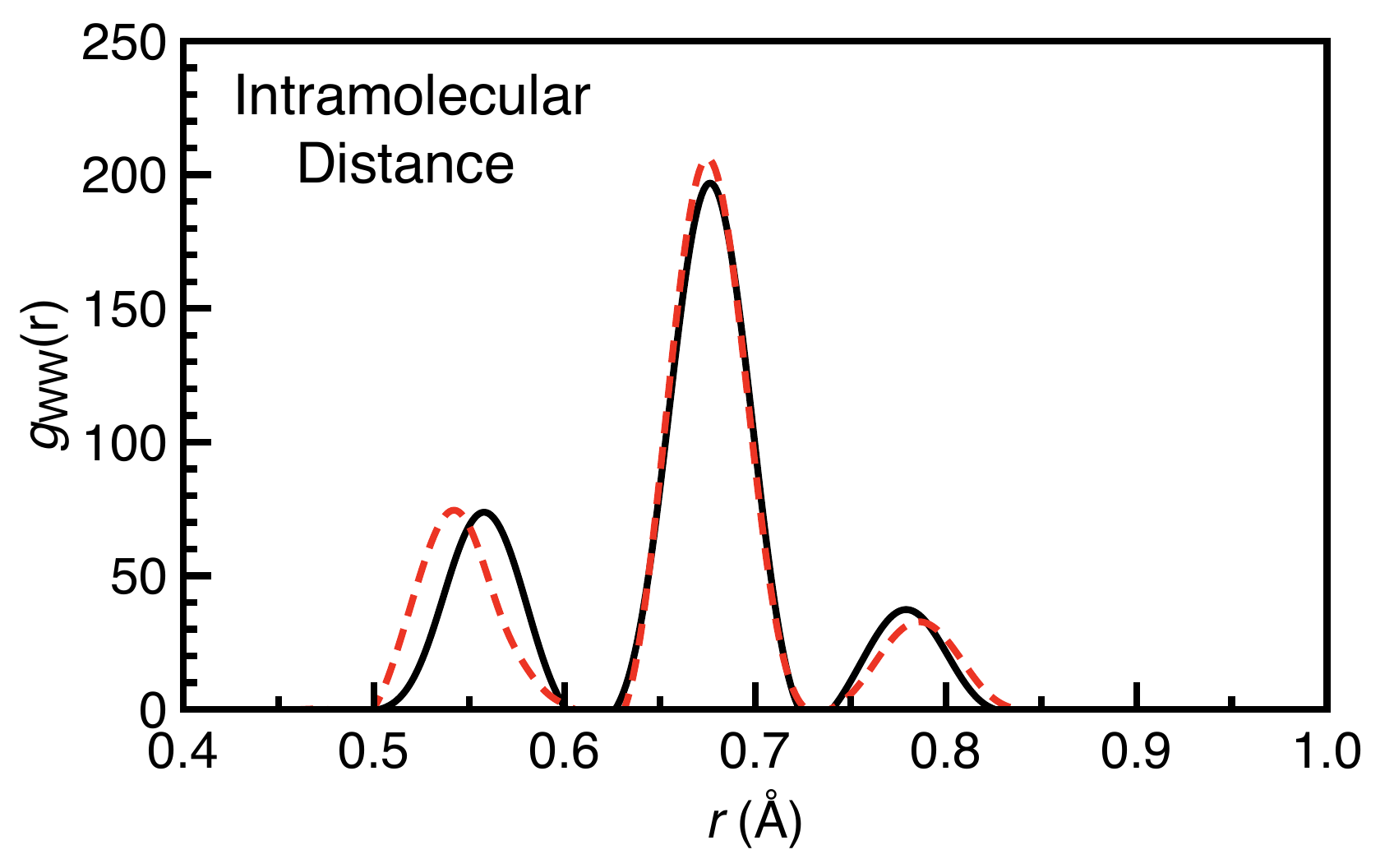}
        \caption{Intramolecular Wannier--Wannier}
        \label{fig:app_intra_xx_50}
    \end{subfigure}

    \caption{ \justifying
    Intermolecular and intramolecular decomposition of the Wannier-resolved RDFs for the $\langle N\rangle\approx50$ QM region.
    }
    \label{fig:app_wannier_decomp_50}
\end{figure}

The decomposition in Fig.~\ref{fig:app_wannier_decomp_50} shows that the largest residual differences occur in the intermolecular channels. The intramolecular peaks remain comparatively localized, whereas the intermolecular oxygen--Wannier and Wannier--Wannier distributions are more strongly influenced by the boundary-dominated environment.

Overall, the 50-molecule results show that the dipole-renormalized reservoir maintains a substantial degree of local molecular and electronic compatibility even in an extreme small-region limit. The disagreement is evident, albeit not dramatic, and it expresses the limitation of a three site model for the MM molecules which can map only the dipole distribution, while for the locality level of a small QM system as this one, the MM environment may need further inputs beyond the dipole, e.g. higher multipole inputs like the quadrupole distributions {\cite{shi2021condensed}}.

\section{Structural properties of the classical reservoirs}
\label{app:bulk_mm_validation}
\begin{figure}[!ht]
    \centering
    \begin{subfigure}[b]{0.49\linewidth}
        \centering
        \includegraphics[width=\linewidth]{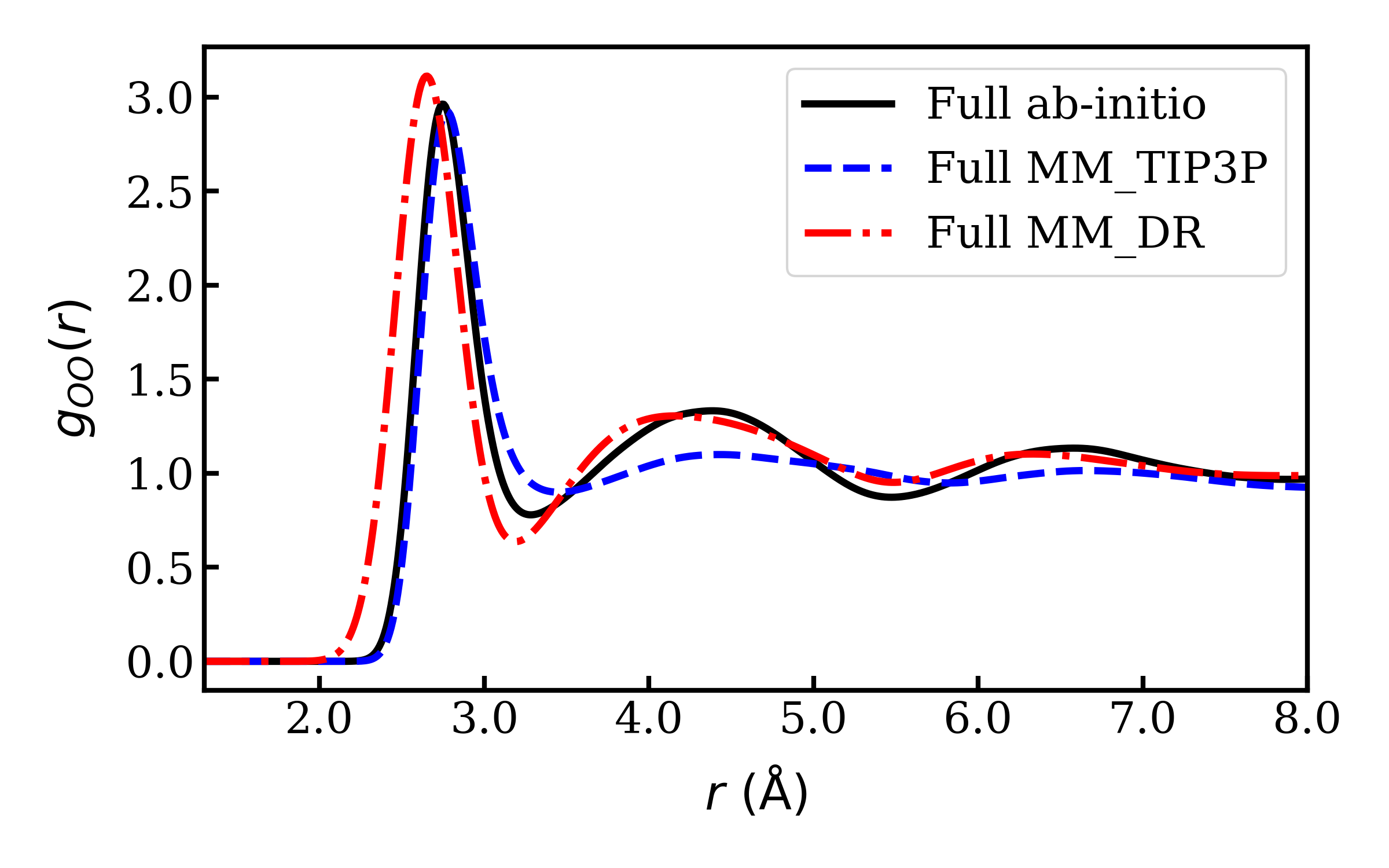}
        \label{fig:bulk_mm_oo}
    \end{subfigure}
    \begin{subfigure}[b]{0.50\linewidth}
        \centering
        \includegraphics[width=\linewidth]{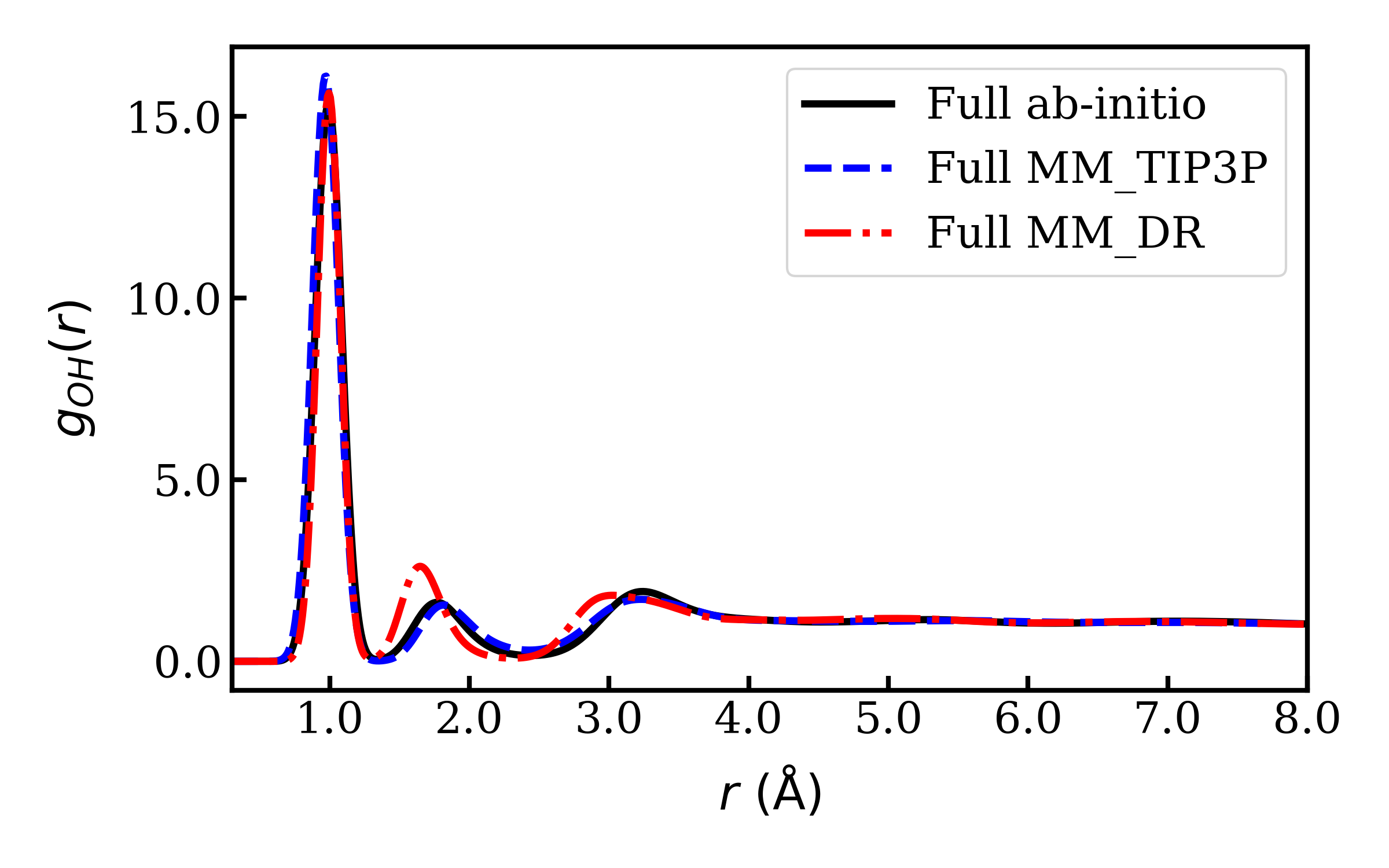}
        \label{fig:bulk_mm_oh}
    \end{subfigure}
    \caption{\justifying
        {Radial distribution functions obtained from full classical MM simulations using the flexible TIP3P and dipole-renormalized (DR) force fields, compared with the full \textit{ab initio} reference: (a) oxygen--oxygen RDF and (b) oxygen--hydrogen RDF.}}
    \label{fig:bulk_mm_rdfs}
\end{figure}

{Before performing the production open-boundary QM/MM simulations, we examined the structural properties of the dipole-renormalized (DR) reservoir in a full classical MM simulation. {Figure~\ref{fig:bulk_mm_rdfs}} compares the oxygen--oxygen and oxygen--hydrogen radial distribution functions obtained using the DR and flexible TIP3P force fields with the corresponding full {{\it ab initio}} reference.}

{As shown in {Fig.~\ref{fig:bulk_mm_rdfs}}, the DR model reproduces the intermediate- and longer-range OO structure more closely than flexible TIP3P, particularly around the second coordination shell. Its first coordination peak is slightly higher and shifted toward shorter distances relative to the {{\it ab initio}} reference. Comparable variations in the position and height of the first OO peak are commonly observed among parameterized water models and reflect quantitative differences in local liquid structure rather than a qualitative change of state {\cite{Wade2018RDF}}. In the present case, the characteristic coordination shells are retained and {$g_{\mathrm{OO}}(r)$} approaches unity at longer distances, confirming preservation of the overall liquid-water structure.}

\begin{acknowledgments}
L.D.S.\ and S.A.\ acknowledge the Nationales Hochleistungs Rechnen alliance (NHR) Project No.\ bec000286 for computational time and the Deutsche Forschungsgemeinschaft (DFG) grant DE 1140/15-1. L.D.S. acknowledge the Deutsche Forschungsgemeinschaft (DFG) through grant CRC 1114 Scaling Cascades in Complex Systems, Project Number 235221301, with  Project C01 ``Adaptive coupling of scales in molecular dynamics and beyond to fluid dynamics''.
\end{acknowledgments}

\bibliographystyle{apsrev4-2}
\bibliography{ref}

\end{document}